\documentstyle[12pt,aaspp4]{article}

\lefthead{Reid}
\righthead{Globular clusters and the thick disk}
\begin {document}
\title{ Hipparcos subdwarf parallaxes - metal-rich clusters and the thick disk}

\author {I. Neill Reid}
\affil {Palomar Observatory, 105-24, California Institute of Technology, Pasadena, CA 91125,
e-mail:  inr@astro.caltech.edu}

\begin {abstract}

We have used main-sequence fitting to calibrate the distances to 
the globular clusters NGC 6397, M5, NGC 288, M71
and 47 Tucanae, matching the cluster photometry
against data for subdwarfs with precise Hipparcos parallax measurements and
accurate abundance determinations. Both the cluster and subdwarf abundance
scales are tied to high-resolution spectroscopic analyses. The distance moduli
that we derive for the five clusters are 12.24, 14.52, 15.00, 13.19 and 13.59
magnitudes, with uncertainties of $\pm 0.15$ magnitudes. As in previous
analyses by Reid (1997) and Gratton et al (1997), these distances are
higher than those derived in pre-Hipparcos investigations.

The calibrated cluster colour-magnitude diagrams provide fiducial
sequences in the (M$_V$, (B$-$V)) plane, outlining the
distribution expected for stars of a particular abundance. We have 
combined the photometric data for NGC 6397 ([Fe/H]=-1.82 dex), M5 (-1.10 dex)
and 47 Tucanae (-0.70 dex) with the mean colour-magnitude relation 
delineated by 
nearby FGK dwarfs to define a reference grid in the (M$_V$, (B$-$V)) plane,
and we have matched this grid against data for stars drawn
from the Lowell proper motion survey, with both Hipparcos astrometry and
abundance determinations by Carney et al (1994). Limiting the
comparison to non-binaries, there are significantly fewer subluminous stars 
than expected given the spectroscopic metallicity distribution. Inverting
the analysis, this implies a reduction by a factor of three in the 
proportion of stars contributing to the metal-poor ([Fe/H] $< -0.4$) 
tail of the Galactic disk. We discuss the implications of these results. 

\end {abstract}


\section {Introduction }

Star clusters, in both the disk and halo of Galaxy, have provided an effective tool for refining
stellar evolutionary theory since the original identification of the main-sequence and
giant branch by Hertzsprung (1905) and Russell (1911).
Given an accurate distance estimate, each cluster provides a
snapshot of the H-R diagram for a given age and metal abundance, mapping not only
variations in the overall morphology of the colour-magnitude diagram, but also 
the evolution of other parameters, such as chromospheric activity, rotation
and binary frequency. Once calibrated, the observed distribution of those
properties amongst field stars can be used to probe the current characteristics and 
the star formation history of the local stellar populations. 

Clusters also serve as chronometers, marking a lower limit to the age both of 
individual stellar populations and of the Galaxy as a whole. Few very old open clusters are
known, but the oldest, Berkeley 17, with an age of $12^{+3}_{-2}$ Gyrs (Phelps, 1997),
provides an estimate of the time since the onset of star formation within the disk. 
This is $\sim2$ Gyrs older than the age derived from analysis of the white dwarf
luminosity function (\cite{os96}).
The lower-abundance globular clusters in the Galactic halo have been identified since
the 1950s  as fossil remnants from the earliest episodes of Galactic star formation, which
occurred during the initial collapse of the protogalaxy (Eggen, Lynden-Bell \& Sandage, 1962). 
Until recently, the 16-20 Gyr age-estimates 
for the most metal-poor of these halo clusters were at odds with cosmologically-based
determinations, which, for conventional cosmologies, indicated ages that were lower by
almost a factor of two (Kennicutt et al, 1996). However, improvements in the theoretical
treatment of convection, helium diffusion and the equation of state, and a more
realistic representation of the population II elemental abundances in stellar models 
have pushed theoretical H-R diagrams to fainter luminosities and redder colours 
for a given age (D'Antona et al, 1997). Equally significantly, new parallax data provided by the 
Hipparcos astrometric satellite (\cite{esa}) have led to accurate distance estimates for a
more extensive sample of nearby halo subdwarfs and, as a consequence, improved
distance estimates to individual globular clusters. As first pointed out by Reid (1997a
- hereinafter paper I), and later corroborated by \cite{g97}, the revised calibration 
leads to larger distances to the most metal-poor clusters, and hence higher luminosities
at the main-sequence turnoff. Combining the theoretical and empirical re-calibrations leads 
to age estimates of between 11 to 13 Gyrs for the oldest halo clusters, significantly 
closer to cosmological determinations.

Once distances have been determined, cluster colour-magnitude diagrams can be used to
investigate the characteristics of the local stellar population, in particular
the abundance distribution. Qualitatively, theoretical models predict that, at 
fixed helium abundance, main-sequence stars of a given mass become hotter (bluer) and brighter 
(at least at visual wavelengths) with decreasing abundance. The quantitative variation in 
colour and luminosity with [Fe/H] is not well-established empirically, 
since, even post-Hipparcos, local subdwarfs with accurate trigonometric parallaxes are 
insufficient in number to provide full coverage of the (M$_V$, colour) plane. 
However, cluster colour-magnitude relations, calibrated by these few fiducial points, 
can trace, for a specific metal abundance, the position of the main-sequence over a 
wide range of stellar masses, given the important provisos that the cluster and
field-star abundances are tied to a consistent observational calibration, and
that there are no systematic difference in secondary parameters, such as the helium abundance.

In this paper we use Hipparcos astrometry of nearby metal-poor stars 
to determine distances to a number of intermediate-abundance globular
clusters, extending to higher abundance the small sample discussed in 
paper I. Taking the
colour-magnitude diagrams of those clusters as references, we deduce the
the likely abundance distribution of local proper-motion stars.
The paper is organised as follows: section 2 discusses the 
calibrating observations, comparing abundance determinations from high-resolution
and low-resolution spectroscopy of both field subdwarfs and globular cluster
giants. Based on this discussion, we define the local subdwarf calibrators; and
consider their distribution in the (M$_V$, (B$-$V)) colour-magnitude diagram.
Section 3 uses these data to estimate distances to intermediate-abundance ([M/H]$>$ -1.3)
globular clusters. Matching the derived absolute magnitude at the main-sequence turnoff against
theoretical predictions, we estimate the age of each cluster. Section 4 discusses
the impact of these new results on the distance scale, particularly on measurements 
of the Solar Radius. The main purpose of the present paper, however, is to
use the calibrated cluster data to probe the characteristics of nearby proper-motion
stars. Thus, in section 5 we use the (M$_V$, (B$-$V)) cluster sequences to define a
reference grid against which we can match the observed distribution of 
proper-motion stars from the Lowell survey (\cite{low}) which have accurate 
Hipparcos parallax measurements. Many of those stars have either spectroscopic
(Carney et al, 1994) or photometric (\cite{sf87}, Schuster et al, 1993) abundance estimates, and
we compare the observed colour-magnitude distribution against that predicted from
previous abundance estimates, paying particular attention to the mildly
metal-poor stars usually associated with the Galactic thick disk.
Section 6 summarises our conclusions.

\section {The subdwarf sample }

\subsection {Absolute magnitudes}

In our initial investigation of the impact of Hipparcos data on the
cluster distance scale (paper I), the calibrating subdwarfs were
drawn from the subset of the Lowell Observatory proper-motion catalogue
which had both Hipparcos astrometry and ground-based
spectroscopic and photometric observations by
\cite{clla94} (hereinafter CLLA). We have since been able to expand our reference 
sample to include a larger number of the classical metal-poor subdwarfs,
and data for the 91 stars which form the basis for the current investigation 
are listed in Table 1. These
stars are chosen as having both high-precision Hipparcos parallax measurements 
($\sigma_\pi \over \pi$) and abundance estimates based on high-resolution
spectroscopy. Where
possible, our photometry is taken from CLLA; if such data are not available,
we have adopted the V magnitude and (B$-$V) colour listed in fields 5 and 37, respectively,
of the Hipparcos catalogue (\cite{esa}). In some cases, the latter colours differ from
the CLLA measurements by several hundredths of a magnitude: for example, HD 19445 is listed 
as (B$-$V)=0.49 in the Hipparcos catalogue, as compared with (B-V)=0.45 in CLLA.
Given the steep slope of the 
main sequence, particularly at M$_V = +5$, these small differences can have
a significant impact on cluster distance estimates derived from main-sequence
fitting.

Table 1 lists the absolute magnitudes given by the Hipparcos
parallax measurements for each subdwarf. As is now well known, systematic biases
can be introduced into a statistical analysis if the calibrating stars
have been selected, either implicitly or explicitly, based on the 
measured parallax, $\pi$. Since the number of stars increases with decreasing parallax, 
observational uncertainties in $\pi$ lead to the mean distance of a given
sample of stars, and hence the mean luminosity, being underestimated. 
Originally quantified for the case of a uniform spatial distribution by \cite{lk73},
\cite{h79} extended the analysis to more general cases, including different spatial
distributions and the effects of introducing a magnitude limit. In paper I
we adopted Lutz-Kelker corrections based on Hanson's formalism, matching the
parallax distribution of the Lowell stars included in the Hipparcos catalogue,
$N(\pi_0) \propto \pi_0^{-3.4}$. There are, however, other factors which should be
taken into consideration, notably the correlation in the Hipparcos data between 
the uncertainty in the measured parallax, $\sigma_\pi$, and the apparent magnitude
(figure 1). The derivation of the individual formal uncertainties is described by 
Arenou et al (1995), who find that systematic errors in the parallax zeropoint 
lie at (or below) the 0.1 milliarcsecond (mas) level. 
The (magnitude, $\sigma_\pi$) correlation is significant 
in the present context, since the calibration stars are selected based on the parallax precision, 
$\sigma_\pi \over \pi$.

We have used Monte Carlo simulations to determine the expected extent of
systematic bias in the current sample, generating a sample of 'stars'
with known properties (M$_V$, $\pi_0$, spatial distribution). Adopting a
particular distribution of observational uncertainties, we can compare the
mean absolute magnitude of a sub-sample, selected based on specific criteria, 
against the known value. The initial calculation simulates the classic
Lutz-Kelker effect. Given an uncertainty of $\sigma_\pi = \pm1$mas
in each parallax measurement, and generating 10$^6$ stars with true distances up to 
1 kiloparsec (i.e. $\pi_0 > 1$ mas), figure 2A shows the difference 
$\Delta M_V \quad = \quad M_V(true) - M_V (obs)$ as a function of parallax precision. 
The turnover in $\Delta M_V$ at lower parallax precision is due to the distance
limit of 1 kiloparsec in the model. We have also plotted the $n=4$ and $n=3$ versions
of Hanson's Lutz-Kelker approximation. As noted by Hanson, the $n=4$ (uniform
distribution) approximation lies slightly below the original Lutz-Kelker calculations.

Figure 2B shows the effect of incorporating a magnitude limit in the $M(\pi_0) \propto \pi_0^{-4}$
simulation. In this case, we have taken M$_V$=5 for all stars, with no dispersion, set
the magnitude limit at V=11.5, and, as in figure 2A, adopted an observational error
distribution of $\sigma_\pi = \pm1$mas.The brighter magnitude limit leads to a
smaller effective distance limit, fewer stars which can be scattered into the sample
from larger distances, and, as a result, $\Delta M_V$ turns over at a higher
parallax precision. 

However, figure 1 shows that one cannot characterise the uncertainties
in the Hipparcos parallaxes by a single value of $\sigma_\pi$. When we adopt a
distribution of $\sigma_\pi$ that is correlated with V magnitude, as in figure 1, the
increased uncertainties in the parallax at fainter magnitudes lead to significantly
more bias in the derived mean absolute magnitudes (figure 2C). Allowing for a dispersion in the
intrinsic absolute magnitudes has little effect on the bias distribution(figure 2D). 
However, the bias is dependent on the effective distance limit, and hence
on the absolute magnitude of the particular set of stars. Figure 2D plots the
$\Delta M_V$ distributions predicted by our simulations for M$_V = 3\pm 0.25$ and $6 \pm 0.25$, 
as well as the M$_V = 5 \pm 0.25$ model. In each case, the observed bias lies close to the 
Hanson $n=3$ approximation, with the bias 'saturating' at the parallax precision which
corresponds to the effective distance limit. Hence, for ${\sigma_\pi \over \pi} > 0.3$, 
the absolute magnitudes for the intrinsically brightest stars, which remain
in the sample at distances of more than 500 pc, are biased to a 
greater extent than the M$_V = 6$ stars. 

The present sample, however, is limited to stars with parallaxes measured to a formal
precision of at least 15 \%. At those high precisions, the predicted bias, at all
M$_V$, is less than 0.15 magnitudes. 
Given the results plotted in figure 2D, we have adopted two conventions in the
main-sequence fitting in the current paper: first, we use Hanson's n=3 approximation
to derive Lutz-Kelker corrections for the reference stars; and, second, we
adopt the Hipparcos parallax measurements directly, applying no LK corrections. 
In most cases, the higher weight accorded to the higher-precision parallax
measurements leads to little difference between the resulting distance estimates.
\footnote{ We should emphasise that the distance moduli derived for
the metal-poor clusters discussed in paper I are little affected  by 
these considerations. Since the final calibration was based on subdwarfs with
parallaxes measured to a precision of 12 \% or better, undertaking main-sequence
fitting without applying any Lutz-Kelker corrections decreases the resultant
best-fit distance moduli by no more than 0.04 magnitude for any cluster.}

\subsection{Abundances and gravities}

Theoretical isochrones predict that the absolute magnitudes and colours of 
main-sequence stars depend on atmospheric composition. Thus, accurate
distance estimates can be derived using main-sequence fitting only if the
globular clusters and the calibrating subdwarfs are tied to a consistent
abundance scale. Abundance measurement in metal-poor stars is complicated by
the fact that oxygen and other $\alpha-$rich elements (Ne, Mg, Si, S, Ca) are 
enhanced by +0.2 to +0.3 dex relative to solar ratios (\cite{sn91}, \cite{sn92})
This is a natural consequence of the dominant role played by type II supernovae
(massive stars) in chemical evolution during the earliest epochs of star 
formation (\cite{mg86}). High-resolution spectroscopy, allowing measurement of
the line-strengths of individual atomic species, therefore constitutes the 
preferred method for abundance determination. 

Until recently, however, conventional high-resolution spectral
analyses have been undertaken for only a restricted number of metal-poor stars. 
Primarily for that reason, in paper I we adopted cluster and subdwarf abundance calibrations 
based on other, more widely-applicable techniques. Extensive
high-resolution data are now available, 
however, through Carretta \& Gratton's (1997) analysis of spectroscopy of red giants 
in 24 clusters, and Gratton et al's (1997a)
re-analysis of high-resolution spectroscopy of nearly 300 field subdwarfs.
We have adopted those calibrations in the present analysis, and here we compare 
the metallicities against our previously-adopted abundance scales. 

\subsubsection {Cluster abundances}

The globular cluster metallicities adopted in paper I are taken from the
\cite{zw85} (ZW85) analysis of large-aperture photometry, where the
latter magnitudes are combined to give a number of reddening-independent
colour indices which effectively measure the colour of the giant branch
and the relative number of blue horizontal branch stars. Both of these
properties are correlated, to first order, with abundance.
The zeropoint of the abundance scale was set by the few clusters which
at that time had high resolution spectroscopic analyses of giant star members.
\cite{cg97} (CG97), on the other 
hand, have analysed high resolution (R$\sim 30,000$) spectra of 160
stars in 24 clusters, using the equivalent widths of individual Fe I
lines to determine mean [Fe/H] values for each cluster. Their sample includes
all of the clusters discussed in section 3, and the same authors have analysed many
of the field subdwarfs in our calibrating sample. An important
qualification is that, while a consistent set of model atmospheres was used in 
the cluster-giant spectral-line analysis, the data themselves are drawn
from a variety of sources and do not constitute an homogeneous sample. 

CG97 compare their abundance determinations against the ZW85 scale (their figure 5).
There is a systematic offset between the two sets of abundance estimates,
with the ZW85 scale giving lower metallicities for [Fe/H] $< -1$. This reflects,
to some extent, the higher solar Fe abundance (by +0.15 dex) adopted as standard 
in the earlier study (see Biemont et al, 1991). The average offset between
the CG97 and ZW85 scales is $\sim0.15$ dex, but for some intermediate-abundance
clusters (such as NGC 288 and M5), the high-resolution abundances are $\sim0.3$ dex 
higher than the old calibration. 

\subsubsection {Subdwarf abundances}

A comparison between recent high-resolution spectral analyses of field subdwarfs
and the results derived from the CLLA survey reveals systematic differences in 
the metallicity scales, comparable to those in the cluster comparison. The CLLA calibration 
is based on echelle spectroscopy, but the practicalities involved in completing a survey 
of over 1000 stars limited most observations to spectra of relatively low
signal-to-noise. Abundances were derived by matching those
data against synthetic spectra spanning a range of effective temperature and
metallicity (\cite{cllk87}), with the resultant metallicities accurate to
0.1 to 0.2 dex. Stars with high-resolution spectroscopic analyses were used to 
define the zeropoint of the abundance scale which therefore, like the
ZW85 cluster scale, is tied to the older solar iron abundance.

We can compare the CLLA results against data derived from two recent,
more conventional analyses of high-resolution spectra: Axer, Fuhrmann \&
Gehren (1994 - AFG) have derived temperatures, gravities and abundances for 115
metal-poor stars, many of which were identified first in the Lowell
survey; and Gratton, Carretta \& Castelli (1997a - GCC) have compiled similar
data for nearly 300 stars, again analysing both their own spectra and literature
equivalent width measurements of Fe I and Fe II lines. 
We have Hipparcos data for sixty-seven
stars from \cite{ax94} and eighty-seven stars from the \cite{gcc97} sample.
The two analyses are based on different model atmospheres, with
GCC using the \cite{k93} calculations, while AFG describe their models as 
almost identical to those of \cite{k79}, and use different analysis techniques.
These differences lead to systematic offsets
in both the effective temperature and the gravity derived for
the 22 stars in common between the two samples (figure 3). The abundances,
however, are in reasonable agreement, with a mean difference of
\begin{equation}
\Delta [Fe/H] \qquad = \qquad [Fe/H]_{GCC} \quad - \quad [Fe/H]_{AFG} \qquad = \qquad 
-0.008 \pm 0.169 
\end{equation}
There are indications of a systematic offset towards negative residuals (i.e.
[Fe/H]$_{AFG}$ higher abundance) for [Fe/H]$_{GCC} < -1.5$.

Comparing both of these high-resolution studies against the CLLA data
shows that the CLLA metallicities tend to be systematically more metal-poor at 
lower abundances (figure 4). This offset must partly reflect differences between
these recent high-resolution analyses and those used to calibrate the CLLA scale.
The discrepancy appears less significant for [Fe/H]$>$-1, but
there are relatively few stars in common at those abundances. Generally, 
the GCC/CLLA residuals show less dispersion, and this may reflect
closer agreement between the temperature scales adopted in those analyses. 
One should also note that the AFG sample includes a larger number of fainter (V $>$ 9)
Lowell proper-motion stars, while the GCC sample consists primarily of well-known
HD subdwarfs. We shall discuss these abundance scales further in section 5. 

In this paper we are concerned mainly with the higher-abundance
([Fe/H] $>$ -1.3) components of the halo and disk. Since there is
reasonable agreement between the AFG and GCC analyses at those
abundances, our sample of local calibrators is drawn from both compilations. 
In cases where a star appears in both samples, we have adopted the 
atmospheric parameters derived by GCC. 
Data for all of the stars included in our final sample of calibrators 
are listed in Table 1. There is considerable overlap between
these stars and the Hipparcos subdwarfs used as calibrators by \cite{g97}.

\subsubsection {Abundances and main-sequence fitting}

Accurate and self-consistent abundance measurements for the clusters
and field stars are required since the validity of the main-sequence 
fitting technique rests on the assumption that one is matching like with
like. Changing either (or both) the cluster or subdwarf abundance scales has
two consequences: first, if one limits the comparison to stars within a
limited abundance range, a given cluster is matched against a different set
of calibrating subdwarfs; second, the $\delta$(B$-$V)$_{[Fe/H]}$ colour corrections 
required to place all subdwarfs on a mono-metallicity sequence are affected.
Either adjustment can lead to a change in distance modulus. In the former
case, the subdwarf sample is small enough that adding or
subtracting one or two stars can change the best-fit distance modulus by
$\sim 0.1 - 0.15$ magnitude.

We can use theoretical isochrones to estimate how differential errors in
[Fe/H], and hence $\delta$(B$-$V)$_{[Fe/H]}$, affect the derived distance
modulus. The \cite{da97} tracks indicate that
${{\partial (B-V)}\over{\partial [Fe/H]}} \sim 0.053$ mag dex$^{-1}$ at 
[Fe/H]$\sim -2$, but ${{\partial (B-V)}\over{\partial [Fe/H]}} \sim 0.143$ mag dex$^{-1}$
at [Fe/H]$\sim -1$. The slope of the lower main sequence is $\sim 5$. Thus,
other considerations being equal, distance determinations to
metal-poor clusters are less sensitive to changes in the relative
abundance scale. A {\sl differential} offset of 0.2 dex between the cluster and
field-subdwarf abundance scales leads to a systematic change in the
distance modulus of 0.14 mag at [Fe/H] = -1 and only 0.05 mag at [Fe/H] = -2.
Unfortunately, the majority of metal-poor calibrators with precise
parallax measurements tend to be closer to the main-sequence turnoff, where
the main-sequence is steeper, compensating to some extent for the reduced
metallicity-dependence of the colours. 

\subsection {The HR diagram for metal-poor stars}

Figure 5a shows the (M$_V$, (B$-$V)) colour-magnitude diagram defined
by the 91 stars listed in Table 1. The stars are divided into six
subgroups by abundance and two reference sequences are plotted:
the mean relation for the nearest stars, derived by \cite{rm93} from
ground-based trigonometric parallax data, and the main-sequence 
relation for the Pleiades, using (V, (B$-$V)) data from 
\cite{mic96} (excluding photometric binaries) and adopting
the Hipparcos-based distance modulus of 5.3 (Mermilliod et al, 1997). 
Figure 5b compares those two sequences against the colour-magnitude
distribution described by all stars in the Hipparcos catalogue with
$\pi > 30$mas and ${\sigma_\pi \over \pi} <0.08$. The 
Reid/Murray relation lies toward the lower-luminosity edge of
the sequence, as one might expect given that the
mean abundance of the local stars is slightly subsolar: 
\cite{wg95} derive $\langle [Fe/H] \rangle = -0.14$ dex from
114 G-dwarfs within 25 parsecs of the Sun. Evolutionary effects also obviously
lead to the stellar distribution being skewed to redder colours for M$_V < 4.5$.
That the Pleiades main-sequence lies at lower luminosities is 
surprising - perhaps the most surprising, and disturbing, result from 
the Hipparcos mission. While one can account for the discrepancy by invoking a 
higher helium abundance (Mermilliod et al, 1997), the required abundance may be as high
as Y=0.35, or nearly 40\% higher than the helium fraction in the Sun and the Hyades
(van Leeuwen \& Hansen-Ruiz, 1997).
Finally, figure 5c plots the (M$_V$, [Fe/H]) distribution for all of the stars in Table 1.

A number of stars in table 1 are identified as binary or multiple
systems, either in the Hipparcos catalogue or by CLLA, 
but there are only six cases where the
companions are both close enough to be unresolved and sufficiently
luminous to affect the photometric properties in a significant manner
\footnote{We note that a number of the `companions' listed in the Hipparcos 
catalogue
are line-of-sight optical companions, rather than gravitationally-bound secondaries.}.
One such star, unfortunately, is BD +62 268, the only extreme metal-poor
star with M$_V > +6$. The other stars are HD 7983, 64606, 111971, BD +46 610
and BD +26 2606 (see notes to Table 1). In partial compensation, we have
been able to add two wide common proper-motion companions to our calibrating
sample: LHS 1279 and HD 158226B. 

Of the latter systems, HD 10607(LHS 1281)/LHS 1279 has received less attention.
Given the high proper motion (0".57 yr$^{-1}$) and small separation (22 arcseconds),
these two stars are very likely to be a physical pair. LHS 1279 does not
have extensive radial velocity observations, and there is therefore no
conclusive evidence on its multiplicity. Ryan (1992) comments that the pair
have discordant photometric parallaxes, but that is on the basis of a linear
(M$_V$, (B-V)) relation and on matching both stars against a subdwarf (M$_V$, (R$-$I)$_K$)
relation. The Hipparcos parallax places HD 10607 (M$_V$=4.06, (V$-$I)=0.67) slightly 
above the (M$_V$, (V$-$I)) relation defined by nearby stars, while, assuming the
same distance, LHS 1279 fall $\sim 0.3$ mag below the relation, at M$_V$=8.17, 
(V$-$I)=1.45. AFG find log(g)=3.98 for HD 10607, so the system may consist of a 
slightly metal-poor subgiant/K-dwarf pair. LHS 1279 has too low an intrinsic
luminosity to serve as a distance calibrator in the present study, but, with HST 
colour-magnitude diagrams extending to close to (even beyond) the hydrogen-burning
limit in some clusters, K and M field subdwarfs, such as Kapteyn's star, LHS 205a
or faint halo subdwarfs discovered using the POSS I/II surveys, 
may well prove useful in future analyses. 

AFG also derive a low gravity for the brighter star, HD 158226 (HIC 85378), in the second
cpm pair. They estimate the abundance as  [Fe/H]=-0.63. 
This star, also known as G181-47, has a lower-luminosity
companion, G181-46, or HIC 85373, which has (within the uncertainties)
an identical parallax and, presumably, an identical abundance. Both stars
lie well above the mean colour-magnitude relation defined by local dwarfs.
(figure 5). G181-46 has a (B$-$V) colour of 0.82 mag and M$_V$=5.2, and 
therefore is clearly an unevolved star, lying within the body of
the main-sequence defined by the Hipparcos data in figure 5b.
Neither component shows velocity variations
characteristic of a spectroscopic binary (CLLA).
It therefore seems likely that the abundance and gravity of HD 158226
have been underestimated, and that both stars are solar-abundance dwarfs.

\cite{ax94} identify a significant fraction of the stars in their sample
as low-gravity subgiants, and comment that failing to allow adequately for the presence of
those stars can bias the conclusions deduced from analysis of large-scale
Galactic structure surveys. However, we have already noted that there
is a systematic offset between the gravities derived in their study and
those calculated by GCC. The Hipparcos parallax data allow us to
place individual stars from the two analyses on the colour-magnitude
diagram. Figure 6 plots the colour-magnitude diagrams described by all of
the AFG and GCC stars with Hipparcos data. We have divided both samples at
log(g) = 4.3, plotting the lower-gravity stars as solid symbols. It is
clear that while most of the stars classed as subgiants by GCC fall in the
expected region of the (M$_V$, (B$-$V)) plane, AFG classify a significant number of stars
with M$_V > +5$ as having low gravity. It is extremely unlikely that
any of these low-luminosity stars have, in fact, evolved onto the subgiant branch.
The hypothesis that a substantial fraction of the Lowell proper-motion
stars are subgiants is therefore not supported by the Hipparcos parallax
measurements. 

There are five stars in the GCC sample which have M$_V \ge 4.75$, but which
are identified as having low gravity. The bluest of these stars is the
metal-poor triple system, BD +26 2606 (log (g) = 4.23), while the four stars
lying close to the nearby-star sequence, HIC 51700, 62207, 84862 and 92532, all
have abundances between -0.38 and -0.51 and gravities in the range 4.1 $<$ log(g) $<$ 4.3.
HIC 84862 is the nearby star, Gliese 672. If any of these four stars are subgiants,
then they are members of a population whose turnoff lies at M$_V \sim$ +5, implying an
age in excess of 20 Gyrs. Lacking any candidate red giant members of such a
population, it seems more likely that all four stars are actually near solar-abundance
disk dwarfs. This highlights the uncertainties involved in basing age estimates for
the local stellar populations on apparently-evolved field stars.

As we noted in paper I, the (M$_V$, (B$-$V)) distribution of metal-poor 
([Fe/H] $< -1$) subdwarfs with M$_V > 5$ can be represented by 
a single main sequence, despite the dispersion in abundance. 
To some extent, this is expected, given the lower sensitivity of (B$-$V)
colours to abundance at [Fe/H]$\sim -2$. However, the small scatter amongst
the [Fe/H]$<-1.3$ stars in figure 5 is still somewhat surprising. We can quantify
this statement by measuring $\delta M_V$, the offset in M$_V$ between the 
observed position of each star and the corresponding point on the nearby-star sequence.
This parameter, subluminosity, can be plotted against abundance and the
behaviour compared with that predicted by theoretical models. We have limited
our analysis to stars with $0.55 \le (B$-$V) \le 0.75$, a colour range that is
sufficiently red that theoretical tracks for different abundances are
almost parallel, even for old, [Fe/H]=-0.7 clusters such as 47 Tucanae. Figure 7
compares the observations against the predictions of the \cite{da97} models. The
latter are computed for a helium abundance of Y=0.23 at metal abundances 
Z$\le 6 \times 10^{-4}$ and
Y=0.235 for higher abundances. While the qualitative agreement is reasonable, the 
data suggest a less rapid decrease in luminosity than predicted for -0.2 $< [Fe/H] < -0.6$
and a plateau in $\delta M_V$ for [Fe/H] $< -1$. 

In section 5 we will use a variant of figure 7 as a means of probing the
abundance distribution amongst Lowell proper-motion stars. This approach 
inverts the technique usually adopted in stellar population studies. Conventionally,
one observes V, (B$-$V) and [Fe/H] and uses those data to infer M$_V$ 
(as in Laird et al, 1998, for example); we have 
observations of V, (B$-$V) and $\pi$ (hence M$_V$) and combine those data to
infer [Fe/H]. Both techniques rely on there being a one-to-one correlation
between position on the HR diagram and [Fe/H]. In both cases, that assumption is
complicated by the likely variation in helium abundance over the lifetime of the Galaxy.
Theoretical models predict that an increased Y leads to lower luminosity at fixed Z,
and it is possible that variations in Y may lead to the increased scatter in
$\delta M_V$ at [Fe/H] $>$-1 in figure 7. Given the difficulties involved in
measuring Y directly in main-sequence stars, the
variation is usually presented as a linear increase with increasing Z, with
the latest estimate (Baglin, 1997) indicating that ${dY \over dZ} \sim 3$.

In terms of calibrating the effects of changing helium abundance, one can 
either incorporate an explicit ${dY \over dZ}$ variation in model isochrone
calculations, or, rely on empirical calibration
and the assumption that the average trend in Y with Z in the calibrators
matches that in the programme stars. The latter approach is usually adopted 
(e.g. \cite{lcl88}), and it is that approach that we follow in section 5. 
One important point should be emphasised: if variations in Y are sufficient 
to invalidate our technique of estimating [Fe/H], then those
same variations also invalidate conventional methods of estimating M$_V$
based on observations of [Fe/H].

\section {Cluster distances by main-sequence fitting}

The main purpose of our compiling astrometric and photometric data for nearby
metal-poor stars is identifying appropriate sets of subdwarfs which can be
used to calibrate the distance moduli of photometrically well-studied
globular cluster systems. In such main-sequence fitting analyses, 
it is customary to emphasise the importance of employing as calibrators
only local subdwarfs whose absolute magnitudes are fainter than
at least M$_V$=5.5. This criterion is justified on the grounds that one
must use stars that are unequivocally on the unevolved main sequence. There
are, however, other concerns which should be borne in mind.

First, if one limits the local calibrators to the lowest luminosity subdwarfs, one is
also forced to match the observed cluster main-sequence at faint {\sl apparent}
magnitudes, where the cluster photometry, particularly the colours, are more liable to
significant, and possibly systematic, observational errors. 
Second, the theoretically-predicted variations in colour due to evolution are 
small - an age difference of 4 Gyrs leads to a change in the (B$-$V) colour at
M$_V = 5$ of barely 0.01 magnitude. The offset between 6 and 10 Gyr
isochrones is even smaller. Those uncertainties in colour are minor
compared with the problems of matching the line-of-sight reddening, defining compatible
cluster and field subdwarf metallicity scales, 
and ensuring that all of the data are on a self-consistent photometric system. 
The steep slope of the main-sequence means that any change in colour 
is amplified by a factor of at least five in the modulus derived by main-sequence 
fitting. Thus, an arbitrary change in
the line-of-sight reddening of only $\pm 0.03$ mag corresponds to 
modifying the best-fit distance modulus by at least 0.1 magnitude.

Finally, increasing age moves a subdwarf to higher luminosities with little
change in effective temperature. Thus, the distance of a cluster can
be overestimated only if the field subdwarfs are, on average, 
significantly {\sl older} than the cluster stars,
while any field stars that are younger than the cluster will tend to bias
the measurement towards too small a distance. Gilmore et al (1995) demonstrate 
that starcount data offer little evidence for field subdwarfs bluer than the 
colour of turnoff stars in globulars, 
suggesting that the majority of the field is at least as old as the
clusters. Setting firm limits on the fraction of the halo that is older
than the oldest clusters is a more difficult observational task, but there is 
currently no evidence for a significant pre-cluster population.

Based on these arguments, we include subdwarfs with M$_V \ge 4.75$
as calibrators in our main-sequence fitting, excluding only stars 
which are unequivocally subgiants. As in paper I, we
calibrate the distances of each cluster by matching the observed 
(V, (B$-$V)) fiducial sequence against subdwarfs of the appropriate abundance,
generally limiting the calibrators to star within $\pm 0.25$ dex of the cluster
abundance. We have used the isochrones calculated by \cite{da97} to adjust the
(B$-$V) colours of the individual subdwarfs to give a mono-metallicity sequence.
In each case, we have adjusted the cluster photometry for the line-of-sight
reddening, so the offset (V$_0$ - M$_V$) gives a direct estimate of the true distance
modulus. 

The formal uncertainties in distances derived through main-sequence
fitting are difficult to quantify in general terms, since the results 
are subject not only to the possible presence of systematic errors in the individual
abundance and reddening estimates, but also depend on the exact distribution
in M$_V$ of the subdwarfs chosen as calibrators for a particular cluster. 
Based on both our own calculations using different subdwarf reference stars
and on an external comparison with the results derived by \cite{g97}, 
we estimate that the distance moduli derived for each cluster are
uncertain by $\pm 0.15$ magnitude. Once uniform high-resolution spectroscopic 
abundance estimates are available for a larger sample of subdwarfs, more 
accurate distance determinations will become possible.

\subsection {The extreme metal-poor clusters}

While the main focus of the present paper is the intermediate and
mildly metal-poor clusters in
the Galactic halo, we discuss briefly how adopting the high-resolution abundance calibration
affects our estimates of the distance moduli of the lower-abundance clusters,
discussed in paper I. There are only two stars in the current sample which are
both unambiguously single and which have abundances [Fe/H] $< -1.7$. These are
HD 19445, to which both \cite{gcc97} and \cite{ax94} assign an abundance of 
[Fe/H]=-1.89 and BD +42$^o$ 3607, at [Fe/H]=-2.01 (\cite{ax94}). Other recent 
high-resolution studies derive lower abundances for the former star: -2.2 dex 
(Magain, 1989); -2.16 (Carney et al, 1997); and -2.0 (R. Peterson, priv. comm.).
One could include higher-abundance subdwarfs as calibrators for these extreme
clusters, but that requires making significant adjustments to the (B$-$V) 
colours. Rather than
apply those corrections, which are ill-defined empirically at present, we
postpone a direct re-analysis of the most extreme metal-poor clusters until
a larger sample of local subdwarfs of comparable abundance have accurate
metallicity determinations. 

We should, however, comment on the recent study by Pont et al (1997), who
have matched the M92 fiducial sequence against seventeen subdwarfs and subgiants
with Hipparcos data, deriving a best-fit distance modulus of (m-M)$_0$=14.68, 
almost identical to the pre-Hipparcos analysis. Their calibrating sample 
includes a larger number of extreme metal-poor ($-1.8 \le [Fe/H] \le -2.6$
on the CLLA scale)
stars than in either of the previous analysis (paper I; Gratton et al, 1997b).
In their main-sequence fitting, Pont et al give equal weight to each star, whereas
we weight the fit linearly by the uncertainty in the parallax. 
Besides applying classical Lutz-Kelker correction, they have used Monte-Carlo
simulations to estimate the biases introduced into a ${\sigma_\pi} \over \pi$-limited
sample by the increase in $\sigma_\pi$ with apparent magnitude (figure 1), and 
the effects of a Malmquist-type bias introduced by observational uncertainties
in the abundance estimates. Since the halo abundance distribution peaks at [Fe/H]$\sim-1.5$,
the latter errors will lead to a systematic {\sl underestimation} of the average
abundance of stars selected as extreme subdwarfs. Given $\sigma_{[Fe/H]} \sim 0.15$dex
(and an intrinsic width, $\sigma_Mv$, of the main-sequence at given [Fe/H]), they calculate
that the latter two effects combine to give a systematic correction to the subdwarf 
absolute magnitudes of +0.06 mag (i.e. opposite to classical Lutz-Kelker) at
[Fe/H]=-2.0 and +0.01 mag at [Fe/H]=-1.0. These corrections are small, and should
be smaller in our current analysis, where the abundances are based on 
high-resolution spectroscopy. More important, the {\sl random} uncertainties in
[Fe/H] within a given abundance scale are clearly smaller than the {\sl systematic}
differences between different scales. 

Pont et al also apply statistical corrections to the absolute magnitudes of the
binaries in their sample, although one should note that their results are not
strongly dependent on the inclusion of those stars. They correct each star by +0.375
magnitudes, based on a comparison with the binary fraction in Praesepe. That
cluster, however, is sufficiently old ($\sim 800$ Myrs; Friel, 1995) that mass 
segregation and evaporation
has enhanced the fraction of equal-mass binaries. Given the small sample size, it 
seems more prudent to simply eliminate known binaries from the calibrating sample,
as in our current analysis. 

\subsection {The [Fe/H]$\sim -1.6$ clusters}

The three intermediate-abundance clusters discussed in paper I 
(M13, NGC 6752 and M5) are served better by 
calibrators with high-resolution spectroscopy, and Table 2 lists the distance
moduli that we derive for these systems. The cluster fiducial sequences are identical
with those used in paper I. Postponing discussion of M5 to the following section, 
our results are in good agreement both with the distances found in our paper I analysis
and by \cite{g97}. The latter is not surprising, given the substantial overlap
between the two calibrating samples. 

We have added to Table 2 the results of our analysis of one additional 
intermediate-abundance cluster - NGC 6397, one of the closest globular clusters.
Conventional estimates of the distance to NGC 6397 place it at a modulus of
between 11.7 and 11.9 magnitudes (Anthony-Twarog, Twarog \& Suntzeff, 1992). 
These estimates, however, are
tied primarily to the distance of one 'classical' subdwarf, HD 64090, whose
Hipparcos parallax is 20 \% smaller than the ground-based measurement cited by
Laird, Carney \& Latham (1988).  The cluster is subject to variable foreground
reddening, but there are no indications of an unusually high value of the 
ratio between total and selective reddening, as is the case with M4 (Liu \& Janes, 1990).
Vandenberg, Bolte \& Stetson (1990 - VBS) estimate a mean differential reddening of
$\delta E_{B-V} = 0.17$mag with respect to M92 (E$_{B-V} = 0.02$) and we
have adopted that value in the present study.

Figure 8a plots the de-reddened fiducial sequence 
listed by Buonnano, Corsi \& Fusi Pecci (1989) for this cluster, matched
against eight subdwarfs from table 1. The best-fit distance modulus is 
12.24 magnitudes, even with the application of no Lutz-Kelker corrections. 
Based on HD 64090 alone, the distance modulus is 12.17 mag. 
CLLA identify that star as a possible single-lined binary, and Pont et al, using
CORAVEL, also suspect the star of low-amplitude velocity variations. However,
eliminating HD 64090 from the distance determination increases (m-M)$_0$ to 
12.26 magnitudes. Matching the NGC 6397 upper main-sequence and subgiant branch to 
an M92 fiducial, VBS derive a difference of 2.17 magnitudes in the apparent distance 
moduli of the two clusters, implying $\delta (m-M)_0 = 2.70$, or (m-M)$_0$ = 14.94 
This takes no account of possible intrinsic differences in the colour-magnitude
diagrams due to the lower abundance of M92. 

The larger distance that we derive for NGC 6397 is
consistent with the \cite{deb95} spectroscopic analysis of
blue horizontal branch stars in the cluster 
(\cite {heb97}). In addition, the observed and predicted magnitudes of
the contact variables V7 and V8, discovered by \cite{k97}, are in
closer agreement for (m-M)$_0$=12.24, although the two SX Phoenicis
stars identified in the same study become significantly more luminous than
expected if either is a cluster member.

\subsection {The [Fe/H]$\sim -1$ clusters M5 and NGC 288}

The recent photometric study by \cite{sb96}, coupled with the
extensive monitoring of the cluster RR Lyrae variables by \cite{re96},
make M5 one of the best-calibrated globular clusters. \cite{zw85} assigned a
metallicity of -1.4 to both this cluster and to the more distant NGC 288, but both
are assigned significantly higher abundances, of -1.1 and -1.07 respectively, 
by CG97. Sneden et al's (1992) analysis of the same M5 red giant data leads to a slightly
lower abundance of [Fe/H] = -1.17 dex. NGC 288 lies towards the South
Galactic Pole, and is subject to very little foreground reddening (\cite{McF83},
\cite{p84}). Bolte (1992) has obtained deep CCD photometry of the cluster, and has used 
those data to determine a fiducial main-sequence extending to $V>22$, (B$-$V)$>$0.9 mag.

Given the similarity in abundance between these two clusters, we follow VBS
and use the offset between the two observed fiducial sequences in (V, (B$-$V)) to
estimate the difference in reddening, deriving 
$\delta$E$_{B-V}$=0.01 magnitude. The line-of-sight reddening towards M5 is generally
estimated as E$_{B-V}$ = 0.02 to 0.03 mag (paper I). Given the low reddening estimates
towards the SGP, we adopt a value of E$_{B-V}=0.01$ for NGC 288 and estimate
E$_{B-V}=0.02$ magnitude for M5. The latter is 0.01 mag lower than the value
used both in paper I and by \cite{g97}. Matching the reddening-corrected main-sequence
against local subdwarfs of the appropriate abundance range gives the best-fit distance 
moduli listed in Table 2. Note that matching the fiducial sequences gives a difference
of $\delta (m-M) = 0.47$ mag., close to the value derived by independent main-sequence 
fitting.

Figure 8b plots the calibrated colour-magnitude diagrams of both clusters together
with the local subdwarf stars. The excellent agreement between the upper
main-sequence, subgiant and giant branches of the two clusters suggests almost identical
ages. The offset of 0.01 to 0.02 magnitude in (B$-$V) at fainter magnitudes
(V $>$ 20.5 in NGC 288) may reflect either an offset in the photometric calibration, 
or a bias in the position of the fiducial sequence introduced by the presence
of significant numbers of binaries in NGC 288 (Bolte, priv. comm.).  This discrepancy
underlines our misgivings about limiting main-sequence fitting to stars on
the lower main-sequence. Our distance modulus of 15.00 to  NGC 288 is $\sim 0.5$
magnitude higher than previous estimates based on pre-Hipparcos parallaxes for subdwarf
calibrators, and $\sim 0.25$ magnitude higher than the value derived by \cite{g97}.
The larger distance is in better accord with spectroscopic 
analysis of blue horizontal branch stars (\cite{heb97}).
Our M5 result is closer to the (m-M)$_0$=14.49
derived by \cite{g97}, and the difference can be attributed primarily to the 
higher reddening adopted in the latter analysis.
Finally, we note that VBS group NGC 288 and M5 together with 
NGC 362 and Palomar 5 in their differential
colour-magnitude diagram analysis. The offsets $\delta$E$_{B-V}$ and
$\delta$V that they derive between NGC 288 and NGC 362 and 
between NGC 362 and Palomar 5 imply distance moduli (m-M)$_0$ of 15.03 
and 17.00 respectively. 

\subsection {The metal-rich clusters 47 Tuc and M71}

47 Tucanae (NGC 104) and M71 are the two best-calibrated metal-rich globular
clusters amongst the systems studied by \cite{cg97}. As with M5, Sneden et al (1994)
derive a slightly lower abundance for M71, [Fe/H]=-0.79 rather than the -0.7
found by CG97. \cite{hes87} undertook
an extensive BV photometric study of 47 Tucanae, estimating the line-of-sight
reddening as E$_{B-V}$=0.04 mag, and we have adopted their derivation of the
fiducial (V, (B$-$V)) relation of the cluster stars. M71 was the subject of a 
detailed investigation by \cite{hod92} who present BV photometry to V$\sim $ 22
magnitude, and they estimate the foreground reddening as a substantial E$_{B-V}$=0.28
magnitude. While the latter factor clearly adds significantly to the uncertainties involved
in main-sequence fitting analysis, a comparative study of the offset between the
M71 and 47 Tuc colour-magnitude diagrams ({\sl a la mode de } VBS)
shows that adopting a differential reddening of $\delta$E$_{B-V}$=0.24 mag
matches closely the colours at the turnoff.

The conventional estimate of the (true) distance modulus of 47 Tucanae is 13.25 mag 
(Hesser et al). \cite{g97} derive a modulus of 13.44 magnitudes, based on five stars with
Hipparcos parallax measurements, an estimated reddening of 0.023 mag  and a cluster
abundance estimate of [Fe/H]=-0.73. Our best-fit modulus is based on nine stars 
and is higher by $\sim 0.13$ magnitude (figure 8c), reflecting primarily the 
the different line-of-sight reddening adopted in our calculation. 
Repeating the analysis with E$_{B-V}$=0.023 mag, we derive
(m-M)$_0$=13.50 mag (no LK corrections). The remaining 0.06 mag. discrepancy rests
with the different sets of subdwarf calibrators used in the two calculations.

\subsection{Cluster ages and the early history of the Galaxy}

The absolute magnitude of stars at the main-sequence turnoff is one of
the most frequently-used age indicators for globular (and, indeed, open)
clusters. Figure 9 compares the empirical results from the present paper
against the (M$_V^{TO}$, abundance) isochrones predicted by the \cite{da97}
models. These models use the recent Rogers, Swenson \& Iglesias (1996)
equation of state; adopt full spectrum turbulence (Mazzitelli et al, 1995),
rather than the mixing length approximation, to deal with convection; and
adopt the bolometric corrections computed by \cite{k93}. 
No oxygen or $\alpha$-element enhancements are included in the DCM models, 
so we follow the \cite{ssc} prescription for adjusting
the cluster abundances to an 'effective' [Fe/H], assuming [$\alpha$/Fe]=+0.3
for [Fe/H] $< -1.2$ and [$\alpha$/Fe]=+0.2 for NGC 288 and M5.

Our distance scale leads to age estimates of no 
more than 11 Gyrs for any of the clusters included in the current sample,
and decreasing the distance modulus by $\sim0.1$ magnitudes, 
matching the Gratton et al results for the metal-poor clusters, increases
the ages to only $\sim 12.5$ Gyrs on average. 
Indeed, even adopting (m-M)$_0 = 14.68$ for M93, the pre-Hipparcos value
and 0.25 magnitudes less than our current estimate, gives an age of only
13 Gyrs for that cluster {\sl calibrated by these isochrones}.
As emphasised in paper I, a crucial aspect of the D'Antona et al isochrones 
is that they use the Kurucz (1993) effective temperature/colour
transformations, leading to a main-sequence $\sim 0.05$ magnitudes bluer
than the corresponding Vandenberg \& Bell (1985) calibration for the same
age and abundance. Most importantly, however, the {\sl shape} of the DCM and Vandenberg isochrones
differ, particularly at extreme abundances ([Fe/H]$\sim -2$), where
matching the lower main-sequence leaves a residual offset of $\sim0.05$ magnitudes
at the turnoff, with the Vandenberg models bluer. Thus, one is inevitably
driven to older ages in analyses based on matching the colours predicted by
the Vandenberg isochrones, and to younger ages if one chooses to use the 
DCM models.

The luminosities predicted by the two sets of models are in closer agreement.
However, the Kurucz scale gives M$_{V \odot} = 4.79$ and BC$_\odot$ = -0.197
mag. (D'Antona et al), as compared with the estimates of M$_{V \odot} = 4.82$ and 
BC$_\odot$ = -0.12 mag derived by Alonso et al (1995). If this difference
of 0.1 magnitude in the inferred M$_{bol \odot}$ can be taken as a simple
zeropoint shift, applicable at all abundances, then the implication is that
age-estimates based on M$_V^{TO}$ in the DCM models will be $\sim 1 - 2$ Gyrs
lower than those derived from the Vandenberg models, which are closer to the
Alonso bolometric scale. These {\sl systematic}
differences in both the underlying theory, and in transforming between the theoretical
and observational planes, are at least as important as the remaining 
discrepancies in the cluster distance determinations in setting the 
uncertainties in the age estimates. 

The two metal-rich clusters, 47 Tuc and M71, merit special mention. Based
on the turnoff luminosities we have inferred for these clusters, both
are younger than the average globular system. This is consistent
with the horizontal branch morphology, with both clusters possessing only
a stubby, red horizontal branch which fails to intersect the instability
strip (although 47 Tuc may have one, unusual RR Lyrae member, \cite{cs93}). 
Theoretical models of HB stars (Dorman, 1992a,b; \cite{lee90}) show that 
the lower the mass of
the star, the higher the temperature achieved on the horizontal branch: that
is, a red HB is characteristic of a higher mass (by $\sim$0.05 M$_\odot$), 
and, presumably, a younger age than the extended blue horizontal branch
which is a feature of clusters like NGC 6397 and NGC 288. The models also
predict that stars on a 47 Tuc-like red HB are $\sim 0.1$ to 0.2 magnitude
more luminous in M$_V$ than the equivalent-abundance, lower-mass stars
which can evolve through the instability strip and become RR Lyraes.

With age estimates for eleven globular clusters, we can combine
these data to map out the onset of star formation within the disk and
halo populations. Figure 10a plots, as a function of abundance, the ages we
deduce for the clusters included in the current paper and in
paper I. We also show the age distribution of Galactic open clusters, 
taking the data from the compilation by \cite{fr95}, together
with the estimated age of the Galactic disk based on the most recent
analysis of the white dwarf luminosity function (\cite{os96}). Figure 10b
shows the corresponding diagram if one adopts the globular cluster
ages calculated by \cite{cds96}, based on distance determined by the
horizontal-branch luminosity calibration, 
\begin{equation}
M_V (HB) \qquad = \qquad 0.79 \qquad + \qquad 0.17 [Fe/H] 
\end{equation}
Chaboyer et al provide age estimates for a variety of different 
horizontal-branch calibrations. The calibration we have chosen gives
one of the youngest age estimates for M92. 

There is an obvious difference in the time spanned by the formation
of the halo clusters in the two scenarios outlined in figure 10, with
the Chaboyer et al (short distance scale) calibration placing
the formation of the oldest clusters (such as M92) over 7 Gyrs
before the formation of younger, intermediate-abundance clusters 
such as M5. More significantly, in the latter scenario, the halo
starts forming stars at least 8 Gyrs before the earliest hint of 
star formation in the disk. \cite{ys79} have shown that there are
feasible mechanisms which can produce a slower overall collapse of the
Galactic halo than the free-fall timescale envisaged originally by
\cite{els}, although even their halo formation occurs over no more
than $\sim 3$ Gyrs. However, their analysis does not provide a 
mechanism for inhibiting formation of the Galactic disk. Given a
crossing time of only a few $\times 10^8$ years, it seems unlikely
that gas clouds can be prevented from colliding and forming stars within
a rotating disk until some 8-10 Gyrs after the first halo clusters 
formed. This is obviously not an issue for the revised globular
cluster timescale, where there is at most a gap of 1.5 Gyrs between
the onset of star formation in the halo and significant star
formation occurring within the Galactic disk.

\section {Cluster RR Lyraes and the distance scale}

\subsection {The distance to the LMC}

One consequence of our revised distance estimates for globular clusters is a 
new calibration of the absolute magnitude of RR Lyrae stars as a function of
abundance. Those variables make an important contribution to the definition of
the extragalactic distance scale by providing an independent calibration of the distance
to the nearer Local Group galaxies, such as the Magellanic Clouds and M31.
In paper I we discussed the impact of the new (M$_V$, [Fe/H]) calibration
on estimates of the distance to the LMC, and noted that the revised
distance modulus of 18.65$\pm 0.1$ (53$\pm 0.7$ kpc) was consistent with the 
recent re-analysis of both the Cepheid distance scale by \cite{fc97}
( (m-M)$_0$=18.7) and the re-calibration of the mira period-luminosity
relation by \cite{vl97a} ( (m-M)$_0$ = 18.54). On the other hand, all of
these results are at odds with the most recent analysis of the expanding
ring in SN1987A by Gould \& Uza (1997), who derive (m-M)$_0 < 18.44 \pm 0.04$.
\footnote{ Note that while Gould \& Uza determine the time of
maximum light as t$_+$=380.7$\pm$6.3 days, from fitting a 
cuspy profile to the N III] emission-line data, the actual 
observed maximum line-flux is $\sim$410 days after the 
supernova implosion. If the latter observation is accurate,
the SN1987A distance estimate should be increased by $\sim 10 \%$
to 52 kpc. ( (m-M)$_0$=18.57). }

As discussed in paper I, Walker (1992a) has compiled data for RR Lyraes
in seven LMC clusters, five of which have abundances close to that of NGC 6397. 
A direct comparison is complicated by the fact that the Galactic cluster 
possesses only blue horizontal branch stars and no RR Lyraes (HB type = 
$(B-R) / (B + V + R) $ = 1). In contrast, the two best-studied LMC clusters, 
NGC 2257 (Walker, 1989) and NGC 1466 (Walker, 1992b) have a horizontal branch
morphology with few red HB stars, but with $\sim 25\%$ of the stars on
the instability strip (HB type $\sim 0.75$). This suggests that the LMC stars are 
on average $\sim 5 \%$ more massive than the NGC 6397 HB stars (\cite{dor92b}), and 
these LMC clusters correspondingly younger. The presence of a more extensive asymptotic giant
branch in NGC 1466 supports this hypothesis.

If we assume that the intrinsic luminosities of the horizontal branch stars in NGC 6397
and the LMC clusters are similar, despite the age difference, then we can estimate
distance moduli for the latter. Walker lists mean magnitudes for the RR Lyraes in
all five [Fe/H]$\sim -1.8$ clusters. The colour-magnitude diagrams for NGC 1466 and
NGC 2257 show that the RR Lyraes are $\sim0.15$ mag fainter than the mean magnitude
of the reddest non-variable stars on the blue horizontal branch.
In NGC 6397, the latter stars stars have (B$-$V)=0.31, $\langle V \rangle$=12.90, or 
(B$-$V)$_0$=0.12, $\langle M_V \rangle$=+0.1, implying that 
[Fe/H]$\sim-1.8$ RR Lyraes have M$_V \sim +0.25$. 
Averaging the resultant distance moduli for the five LMC clusters 
gives $\langle ({\rm m-M}_0 \rangle = 18.71 \pm 0.06$ mag.
This calculation makes no allowance for possible effects due to the depth of the
LMC, but if we follow Walker and assume that the clusters lie in the plane defined by
the HI disk, then the mean modulus remains unchanged at 
$\langle ({\rm m-M}_0 \rangle = 18.71 \pm 0.07$ mag. Given the morphological differences 
between NGC 6397 and the LMC clusters, the uncertainty 
in this estimate is at least $\pm 0.15$ magnitude.
However, the derived distance is more consistent with the longer distance 
scale than with the SN1987A analysis. 

\subsection {The distance to the Galactic Centre}

In paper I we noted that the absolute magnitude derived for the M5 variables
was in poor agreement with results derived from statistical parallax analysis 
of Solar Neighbourhood RR Lyraes (\cite{l96}). Since our cluster abundances were
based on the ZW85 scale, the poor agreement might be interpreted as due to a
significant difference between the metallicity of M5 and the mean abundance of even
the halo field-star sample in the latter analysis. Under the CG97 abundance 
scale, however, M5 has [Fe/H]=-1.1, and our distance modulus of 14.52$\pm$0.1 
(E$_{B-V}=0.02$) implies
M$_V$=0.5$\pm$0.1 for the cluster RR Lyraes, $\sim 0.3$ mag brighter than the 
results derived for field RR Lyraes of the same abundance, and $\sim 0.15$ mag brighter
than the absolute magnitudes predicted for [Fe/H]=-1 horizontal branch stars
by Caloi et al's (1997) theoretical models. Moreover, under the CG97 calibration
M5 has an abundance within 0.1 dex of the average metallicity of
the RR Lyraes in the Galactic bulge (\cite{wt91}). Therefore re-calibrating the M5 
variables affects directly one method of estimating the distance to
the Galactic Centre. Before considering how our current results impinge on this
analysis, we review briefly other methods of estimating this quantity.

\subsubsection{The Solar Radius}

Reid (1993) has reviewed the various techniques used to determine R$_0$, 
combining different measurements to derive a ``best'' estimate of R$_0$=8.0$\pm$0.5 
kpc, while Huterer et al (1995) provide a summary of some more recent results.
The most direct methods rely on
geometry, either direct measurement of the proper motion of the radio
source at the Galactic Centre, or the assumption of uniform expansion of water
masers. Thus, observations of the 
H$_2$O masers in the Sgr B2 (North) source, which lies $\sim 300$pc from the Galactic
Centre (\cite{r88}), indicate R$_0$=7.1 $\pm 1.5$ kpc. Similar analysis of the H$_2$O 
masers in W49 leads to R$_0$=8.1 $\pm 1.1$ kpc (Gwinn et al (1992)).

Direct measurement of the proper motion of Sgr A* yields values of
($\mu_l, \mu_b$) = ($-6.55\pm0.34, -0.48\pm0.12$) mas yr$^{-1}$,
where the error-bars give the 2$\sigma$ uncertainties (Backer, 1996)
Given that this proper motion reflects only the reflex solar motion, then 
adopting a local rotational velocity of $\Theta_0 = 220 \pm 20 {\rm km s}^{-1}$ 
(Brand \& Blitz, 1992) and a Solar motion of V$_\odot = 12$ kms$^{-1}$
with respect to the Local Standard of Rest implies R$_0$=7.5$\pm 0.7$ kpc,
or (m-M)$_0$=14.38$^{+0.19}_{-0.21}$ magnitudes.

Caldwell \& Coulson (1987) matched an axisymmetric Galactic rotation model
against photometry and radial velocity measurements of field Cepheids to 
derive an estimate of R$_0 = 7.8 \pm 0.7$ kpc. Huterer et al (1995)
describe other kinematic approaches, combining radial velocity and proper-motion
measurements of Bulge giants, but point out that the existence of a 
bar (Blitz \& Spergel, 1991) may well vitiate the assumption of isotropic
velocity dispersions which underlies those analyses.
Finally, given a flat rotation curve locally, R$_0$ can be derived from 
the Oort constants, but current determinations of the latter quantities are
sufficiently uncertain to provide only poor constraints.

\subsubsection {RR Lyraes and the distance to the Galactic centre}

Oort \& Plaut (1975) carried out the first extensive survey of RR Lyraes
in the Galactic Bulge, analysing photographic data for over 1200 variables in
several fields to derive R$_0 = 8.\pm0.6$ kpc. Most surveys since then
have concentrated on the lightly-reddened Baade's window (see Carney et al, 1995, 
for a summary). In particular, Walker \& Terndrup (1991) combined 
spectroscopic and visual data for 59 RR Lyraes to estimate a Solar Radius of 
R$_0 = 8.2$ kpc for a reddening law where the ratio of total to selective
extinction, R, is 3.1; and R$_0 = 7.7$ kpc for R=3.35. Those distances are
based on a mean RR Lyrae absolute magnitude of $\langle M_V \rangle = 0.85$, consistent
with the Carney et al (1992 - CSJ) Baade-Wesselink scale. Adopting 
$\langle M_V \rangle = 0.50$, as indicated by our M5 results, increases the inferred 
values of R$_0$ to 9.6 and 9.0 kpc respectively. 

Interstellar reddening is much less of a problem at near-infrared 
wavelengths, and Fernley et al (1987) demonstrated that there appeared
to be little metallicity dependence in the (M$_K$, log(period)) relation.
Building on this, Carney et al (1995) used infrared observations of
Galactic Centre RR Lyraes to estimate R$_0 = 7.8$ kpc for an absolute
calibration tied to the CSJ calibration. They also calculate results for
a relation of the same slope, but scaled to match an LMC distance modulus 
of 18.5 mag, a zeropoint correction of -0.3 mag. Feast (1997) has argued
that the latter adjustment is incorrect, since it is based on a
biased derivation of the RR Lyrae (M$_V$, [Fe/H]) relation.\footnote
{Note, however, that Feast \& Catchpole's (1997) re-calibration of
the LMC distance based on Hipparcos results leads to a distance modulus of
18.7, in good agreement with our RR Lyrae calculations, and a consequent
offset of -0.2 mag. in the absolute magnitudes of the LMC RR Lyraes plotted
in Feast's figure 1.} Feast derives an (M$_K$, log(P)) relation, based on
Fernley's (1994) Baade-Wesselink results for field RR Lyraes, of
\begin{displaymath}
M_K \qquad = \qquad -2.566 \quad log P_0 \quad - \quad 1.034
\end{displaymath}
Combined with the Carney et al data on the Bulge RR Lyraes, this leads to 
an estimate of R$_0 = 8.1 \pm 0.4$ kpc. 

Longmore et al (1990) present K-band observations of eighteen RR Lyraes in
M5. Calibrating their absolute magnitudes using our estimate of the
distance modulus (14.52 mag), we derive a best-fit period-luminosity relation of
\begin{displaymath}
M_K \qquad = \qquad -2.566 \quad log P_0 \quad - \quad 1.34
\end{displaymath}
where we have fixed the slope at the value derived by Feast, and solved
only for the zeropoint. This, in turn, implies R$_0 = 9.3 \pm 0.7$ kpc,
or (m-M)$^{GC}_0 = 14.84$. 

\subsubsection {Summary}

The preceeding discussion leads to an apparent paradox. If the Galactic Centre lies
at a distance R$_0 = 8.0$kpc, then the Bulge RR Lyraes imply (m-M)$_0 = 14.19$
to M5 and $\langle M_V (RR) \rangle = 0.85$ at [Fe/H]=$-1.1$. This is
consistent with $\langle M_V (RR) \rangle$ deduced from both Baade-Wesselink (CSJ)
and statistical parallax (Layden et al, 1996) analyses of field RR Lyraes, 
but disagrees by 0.35 magnitudes with the absolute magnitude of the M5 stars derived 
on the basis of (m-M)$_0 = 14.52$. 

On the other hand, the Galactic Centre-based distance to M5 of 6.9 kpc.
is significantly shorter than that derived by other studies.
Sandquist et al (1996) use pre-Hipparcos subdwarf data to derive (m-M)$_V = 14.41$ 
to M5 for [Fe/H]=-1.4 (increasing the cluster abundance increases the inferred distance
modulus), while matching their fiducial sequence to
Vandenberg's enhanced-[$\alpha$/Fe] models ([Fe/H]=-1.3) gives (m-M)$_V = 14.50$; and
HD 103095 alone, ([Fe/H]=-1.22, M$_V=6.61\pm0.0155$, (B-V)$_{[Fe/H]=-1.1}=0.76$),
the pivotal subdwarf in pre-Hipparcos cluster main-sequence fitting, leads to 
a true distance modulus of 14.43 mag (E$_{B-V}$=0.02) and $\langle M_V (RR) \rangle = 0.57 \pm 0.1$,
and (m-M)$^{GC}_0 \sim 14.75$ mag.
\footnote{Note that all of these distance estimates are consistent at the 2$\sigma$ level with
distance moduli of (m-M)$_0 \sim 14.3$ for M5 and R$_0 \sim 8.4$ kpc.}

The paradox exists, however, only if the globular cluster RR Lyraes are
identical to both the local field stars and the Galactic Centre variables. 
That assumption underpins the use of these stars as distance indicators, but
is not inviolable. Dorman's (1992a) models show that the absolute magnitude 
of the horizontal branch depends on the CNO abundance, with increased abundance 
leading to higher luminosity. However, even enhancing the CNO fraction by a factor of 60 
leads to $\Delta M_{bol} \sim 1.0$ magnitudes, and there is no
evidence for such substantial variations amongst either Bulge stars (McWilliam \&
Rich, 1994) or local RR Lyraes (Clementini et al, 1995). Indeed, the latter
analyses suggest [O/Fe]$\sim +0.7$, higher than that inferred for cluster stars, while
a comparison of the ($\Delta$S, [Fe/H]) relation for field and cluster RR Lyraes 
suggests that the latter stars have a higher $\Delta$S by $\sim$1 at [Fe/H=$-2$,
suggesting weaker calcium lines and lower [$\alpha$/Fe] (Clementini et al, 1995; CG97).
Both differences would suggest a luminosity variation in the opposite sense, 
i.e. M$_V$(cluster)$>$M$_V$(field).

However, there is another possible source of luminosity variation. 
High-resolution spectroscopy of globular cluster giants, particularly by
the Lick-Texas group (Sneden et al, 1994 and refs within) has revealed star-to-star
variations in C, N, O, Na, Mg and Al abundances which are both correlated
(sometimes anticorrelated, e.g. Na/O) and tend to become more
anomalous with increasing luminosity. As summarised by Shetrone (1997), the latter
characteristics suggest strongly that mixing within the convective envelope
(perhaps drive by rotation) and dredge-up of nucleosynthesis products is
the dominant process governing this behaviour, although primordial variations may
also contribute in some clusters (e.g. $\omega$ Cen). Significantly, neither
the Na/O anticorrelation nor variation in [Al/Fe] and [O/Fe] has been detected 
amongst field halo giants (Shetrone, 1997).

Sweigart (1997) has pointed out that if mixing extends to sufficient depth within
the envelope to dredge up aluminium, then helium is also dredged up, increasing
the helium fraction within the hydrogen envelope. This has important consequences for
post-RGB evolution, notably an increased luminosity on the horizontal branch.\footnote{The
absolute magnitude at the tip of the RGB also increases} While Sweigart's models
are limited to a single abundance (Z=0.0005, [M/H]=$-1.6$), he estimates the latter variations
as $\Delta log(L) \quad \sim \quad 1.5 \Delta X_{mix}$, where $\Delta X_{mix}$
is the difference in the hydrogen abundance within the envelope and at the base of
the dredge-up region. Sweigart suggests that this mechanism could produce 
the unusually bright RR Lyraes present in some clusters (such as 47 Tuc), while
deeper mixing in metal-poor cluster
giants might account for both the period-shift effect (Sandage, 1982; 1993a)
and the steep (M$_V$, [Fe/H]) relation deduced by Sandage (1982, 1993b; and 
in our paper I) for cluster RR Lyraes. 
Moreover, a systematic difference of $\sim 8\%$ in $\Delta X_{mix}$ 
between the M5 stars and the average value in the field could account for 
$\Delta M_V \sim 0.3$ mag. 

Systematic cluster/field differences in horizontal branch absolute magnitudes are
not implausible, given the evident scarcity of deep mixing amongst field halo giants,
and might stem from environmental (density-related?) variations in the respective
star formation sites. Clearly, such variations would have serious consequences for the
utility of RR Lyraes as standard candles, although those uncertainties might
be minimised by restricting the comparison to variables drawn from similar (present-day)
environments, as in the Galactic/LMC cluster variable comparisons in section 5.1. 
Regardless of the distance scale complications, deep mixing provides a mechanism
which might reconcile the independently well-established $\langle M_V (RR) \rangle$
estimates for the [Fe/H]=$-1.1$ RR Lyraes in M5 and in the disk and Bulge.

\section {Probing the abundance distribution of the thick disk}

The Hipparcos subdwarf parallax data permit distance determination for
globular clusters spanning an abundance range of over 
1 dex, from 47 Tucanae at -0.7 dex to NGC 6397 at -1.82 dex. Those clusters define
a series of fiducial sequences in the (M$_V$, (B$-$V)) plane which map
the location of stars at a number of specific metallicities. Those sequences
therefore define a reference grid which can be used to calibrate the
relation between subluminosity, $\delta M_V$ (as defined in section 2.3), 
and abundance. Figure 11 shows the correlation between subluminosity and 
abundance for (B$-$V) colours of 0.55, 0.65 and 0.75 magnitude. This is 
analagous to the variation
plotted in figure 7, save that the points at lower abundance are defined by
the cluster sequences, rather than by individual stars. Since the former are
calibrated by the latter, it is not surprising that the behaviour of $\delta M_V$
in the two diagrams is very similar. Given the (M$_V$, (B-V)) distribution of
the $\pi > 30$ mas Hipparcos stars plotted in figure 5, we have placed the
solar metallicity point 0.25 mag above the nearby-star sequence.
Our aim is to use the mean relation defined in
this figure as an alternative method of examining the abundance distribution of 
stars in the Lowell proper motion survey, the higher velocity stars in the Solar 
Neighbourhood.

Based on the dispersion of points in figure 7, we estimate random uncertainties
of 0.2 to 0.3 dex in our metallicity estimates. The systematic accuracy of
the technique depends on how well the calibrating cluster sequences
match the proper motion stars. Of the three fiducial clusters, 47 Tuc is the
most important, since we are interested primarily in studying the mildly
metal-poor stars in the Galactic disk. Eggen (1990) has pointed out that
47 Tuc giants follow the (CN-strength, T$_{eff}$) relation 
(traced by the DDO indices C$_m$ and (42-48)) defined by old disk stars
in the Arcturus group, rather than matching the behaviour of members
of more metal-poor halo clusters. Moreover, Sneden et al (1994) find
that Arcturus itself has a pattern of [$\alpha$/Fe] abundance enhancements which
closely matches that observed in giants of the 47-Tuc-like cluster, M71. Given
these similarities, it is reasonable to take 47 Tuc as representative of the
mildly metal-poor old disk. One possible difference might be the helium
abundance: if a significant fraction of the field stars with [Fe/H]$\sim -0.5$
are younger than 47 Tuc, one might expect higher Y, and a lower luminosity
than predicted by the cluster calibrators. In that case, our $\delta M_V$
calibration will underestimate the metallicity of those stars - i.e. we will
overestimate the fraction of mildly metal-poor stars in the sample.

The cluster distance-scale calibration rests on a relatively small number of 
stars with well-determined abundances. However, we have Hipparcos astrometry
for some 2400 stars drawn from the Lowell proper motion survey, with 
nearly 1700 having parallaxes measured to a precision of better
than 10 \%. Since the stars are selected from a proper motion sample, the
contribution made by higher velocity stars in the Solar Neighbourhood is
amplified, and the sample should be well-suited to probing the nature of the 
thick disk. Before describing our analysis, we summarise the results of previous 
studies of that population.

\subsection{The properties of the thick disk}

The thick disk population was identified originally by \cite{gr83} through
analysis of starcount data towards the South Galactic Pole. The number density
of stars between 1 and 5 kpc. above the Plane is significantly higher than was expected
on the basis of the then-standard old disk/halo galaxy models, and Gilmore \& Reid
identified the excess stars with a population having a scaleheight of $\sim 1500$ pc
and local density $\sim 2 \%$ of the old disk. More recent analyses, while
confirming the complex nature of $\rho$(z) (where z is height above the Plane), favour
a higher local normalisation for the extended component, with values in the range
$\sim 5$ to 10 \% $\rho_0$(disk) (\cite{ro96}; \cite{red96}). The associated density law 
(exponential or sech$^2$) is correspondingly steeper to accommodate the observed number densities
of stars at moderate heights above the Plane, with typical estimates of an
exponential (or equivalent) scaleheight of $\sim$800 parsecs. 
The relatively
red colour of the turnoff at z$\sim1$ kpc indicates that this is an
old population, formed either during the initial stages of dissipational collapse or perhaps as
a response to an early merger with a moderately massive companion (\cite{ro96}). 
There is, however, no compelling evidence at present to decide whether this component
is a separate population, or a subset of the Galactic old disk.

There is also some disagreement over the associated properties - abundance, 
kinematics - of the thick disk, reflecting the difficulty of unambiguous segregation
of those stars from members of the old disk. Most studies have converged
on average abundances of [Fe/H] = -0.5 to -0.7 dex (\cite{maj93} and refs within;
\cite{gil95}; \cite{ro96}), with a significant low-metallicity tail extending
to at least [Fe/H] $\sim -1.6$ (\cite{mor93}; \cite{maj93}; \cite{bsl95}). On
the other hand, 
\cite{red97} find that spectroscopy of M-dwarfs at $\sim 2$ kpc above
the Plane (i.e. where the thick disk is expected to be dominant)
suggests an average abundance closer to solar values. In
analyses particularly relevant to the present study, 
\cite{ns91} and \cite{s93} have used accurate Stromgren photometry to
derive abundances for over 1200 stars, also drawn primarily (via \cite{sf87}) from the
Lowell survey. They argue that the abundance distribution is
well-matched by adding a thick disk component with $\langle [Fe/H] \rangle = -0.5 \pm 0.1$,
$\sigma_{[Fe/H]} = 0.25$ dex to the old disk ($\langle [Fe/H] \rangle = -0.11, 
\sigma_{[Fe/H]} = 0.16$) and halo ($\langle [Fe/H] \rangle = -1.30, 
\sigma_{[Fe/H]} = 0.60$) components. None of these studies find evidence for a 
significant abundance gradient with height above the Plane. 

Kinematically, early studies favoured a population with a substantial
($\sim 100$ km s$^{-1}$) rotational lag with respect to the old disk
(\cite{gr83}; \cite{wg86}), although this hypothesis was based more on 
identifying the local metal-rich RR Lyraes as thick disk tracers 
(based on the inferred $\sigma_W$) than on any direct observations.
More recent investigations find a local asymmetric drift of $\sim 30$ kms$^{-1}$
(\cite{maj93}; \cite{ro96}), with velocity dispersions significantly higher
than those of the old disk. \cite{s93}, in particular, propose that the
velocity distribution of stars in their sample is well-matched by a three Gaussian
distributions, with the thick disk contributing stars with $\langle V_{rot} \rangle = 180 \pm 50
{\rm km s^{-1}}$ (i.e. V = -40 kms$^{-1}$), $\sigma_V$ = 50 kms$^{-1}$ and 
$\sigma_W$ = 46 kms$^{-1}$. The old disk and halo are characterised as respectively
$\langle V_{rot} \rangle = 205 {\rm km s^{-1}}$, $\sigma_V$ = 20 kms$^{-1}$ and 
$\langle V_{rot} \rangle = 20 {\rm km s^{-1}}$, $\sigma_V$ = 100 kms$^{-1}$, 
$\sigma_W$ = 86 kms$^{-1}$. On the other hand, one should note that the
last-mentioned analysis does not make quantitative allowance for selection effects in
the sample definition, notably the bias toward higher velocity stars implicit in
any proper-motion star sample. Thus, while these data provide evidence that a
simple two-component Gaussian (old disk/halo) is not a good match to the local
velocity distribution, their model with three {\sl separate} components is consistent 
with, but not required by, the observations.

Nonetheless, characterising the old disk and thick disk components as
kinematically-separable is a useful technique for estimating the completeness
of the subset of Lowell proper-motion stars which form the current Hipparcos sample. 
As described below, the Hipparcos stars were selected from a magnitude-limited 
subset of the Lowell catalogue, V $\le 11$. However, these stars do not necessarily 
constitute a magnitude-limited sample - the higher-luminosity stars in the disk
populations form proper-motion-limited samples. 

In a proper-motion limited sample, the distance limit, and hence the volume surveyed, 
is set by the average tangential motion. Hence, 
fractional contribution from a higher-velocity
population can be amplified. The amplification factor is reduced if the distance limit
is imposed by other factors, such as apparent magnitude or, since we limit our subsequent
analysis to stars with ${\sigma_\pi \over \pi} < 0.15$, 
parallax precision. In the case of the old disk, the velocity ellipsoid is
\begin{equation}
 U_0 = -9, V_0 = -22, W_0 = -7; \sigma_U = 43, \sigma_V = 31, \sigma_W = 25 {\rm kms}^{-1} 
\end{equation}
(\cite{red95}), implying an average tangential velocity of $\langle V_T \rangle$ = 77  kms$^{-1}$.
For a proper motion limit of $\mu > 0".26 {\rm yr}^{-1}$ (as in the Lowell survey) 
this corresponds to
a median distance of r$_m$ $\sim $ 47 pc, while $\sim 80 \%$ of the sample should
lie within 67 pc ( (m-M)$\sim$4.1 mag). The remaining 20\% of the stars with $\mu > 0".26
{\rm yr}^{-1}$ lie at distances extending beyond 120 parsecs. Thus, unlike a magnitude-
or volume-limited sample, the stellar distribution in a proper-motion-defined sample 
does not peak at the apparent magnitude limit. For present purposes, we take the
80\%-complete sampling distance, r$_{80}$, as the effective distance limit.
Since the Hipparcos/Lowell dataset has an {\sl a priori} apparent magnitude limit 
of V$ \sim 11.0$, these simulations show that, for disk stars, the transition between a 
proper-motion limited and magnitude-limited sample occurs at M$_V \sim 7.0$. 

For the higher-velocity component, \cite{bsl95} have used their observations of
spectroscopically-selected metal-poor star to estimate the velocity ellipsoid
of thick disk stars as
\begin{equation}
 U_0 = 0, V_0 = -25, W_0 = 0; \sigma_U = 63, \sigma_V = 42, \sigma_W = 38{\rm kms}^{-1} 
\end{equation}
values comparable with those derived by \cite{s93}. These motions correspond
to an average tangential velocity of $\sim 107$ kms$^{-1}$, 
r$_m \sim $ 66 pc and r$_{80} \sim 94$ pc ((m-M)$\sim$5 mag. For
an apparent magnitude limit of V$\sim 11.0$, these kinematics imply that the
sample is proper-motion limited for M$_V < 6.0$ and magnitude-limited
for lower intrinsic luminosities. Within the proper-motion limited r\'egime, 
the relative number of thick disk to old disk stars in the proper-motion catalogue
(given by the ratios of the cube of the tangential velocities) 
is amplified by a factor of $\sim 2.7$ over the local space densities.
Finally, note that only 20 \% of the thick disk stars within the volume 
defined by r$<$94 pc are included in the proper-motion sample.

In contrast, the halo velocity ellipsoid
\begin{equation}
U_0 = 0, V_0 = -236, W_0 = 0; \sigma_U = 153, \sigma_V = 93, \sigma_W = 107{\rm kms}^{-1}
\end{equation}
derived by \cite{bsl95} implies $\langle V_T \rangle = 340 $ kms$^{-1}$. Analysis
of the LHS catalogue (\cite{re97b}) suggest that the actual average tangential velocity is
$\sim 20\%$ lower, but even in that case, r$_m \sim 180$ pc, and the effective
distance limit in the present analysis is set by parallax precision in the Hipparcos
dataset. As a result, halo subdwarfs make a small contribution to the
final sample discussed here.

\subsection{Proper motion samples, Hipparcos  and selection effects}

The main aim of the current section is to use the Hipparcos parallax
measurements to probe the (M$_V$, (B-V)) distribution of the stars
identified by CLLA as members of the Galactic thick disk. It is
therefore essential that we establish that the selection effects which
are inherent in the definition of the Hipparcos sample do not lead to
any systematic bias in our analysis of those stars.

 First, both the CLLA sample and the subset of the Hipparcos catalogue
discussed here are drawn from the Lowell proper motion survey. The CLLA
sample was chosen based on the Lowell colour class (classes 0, +1),
supplemented by spectral type information from Luyten's catalogues, with
the aim of identifying F and G stars. (We should note that the Hipparcos
sample includes $\sim300$ stars with (B-V)$<$0.75 which are not included
in the CLLA sample, most of which are listed as colour class +2 in the Lowell 
catalogue). The effective magnitude limit is V$\sim$14.5, 
although a significant number of fainter stars are included in the sample.
There are 1464 stars with UBV data, 1261 of which have spectroscopic
abundance estimates. 

The Hipparcos dataset of 2400 stars was selected from the $\sim 3100$ with
m$_{pg} \le 13$ (V $\le 11.5$ for non-degenerates), with the selection based 
on the scheduling constraints of 
the astrometric mission, notably meeting the surface density limit of 3 to 4 stars
per square degree. The longer observation time required to obtain accurate
data for fainter stars also resulted in the final sample including $\sim 80 \%$ 
of Lowell catalogue to V=10, but only $\sim 50 \%$ of the stars between
10th and 11th magnitude. However, the latter restrictions were applied {\sl after}
the V $<$ 11.5 sample was defined, and none of the stars was selected for
observation based on foreknowledge of parallax, absolute magnitude or abundance.
As a result, the stars observed by Hipparcos represent a randomly-selected subset 
of the Lowell catalogue. Our aim is to define an unbiased sample to match against
the CLLA result, and that aim can be achieved by limiting analysis to stars
in the proper-motion-limited r\'egime.

Based on the discussion in the previous sub-section, we expect 
the Lowell/Hipparcos dataset to be proper-motion limited for disk/thick disk
dwarfs with M$_V < 6.0$. The overwhelming majority of thick disk stars have
[Fe/H] $> -1$, so this absolute magnitude limit corresponds to (B-V)$< 0.70$.
We can test whether our expectations are reasonable
by considering the apparent magnitude distribution of stars in the CLLA
sample with 0.55 $\le$(B-V)$\le$0.70 (imposing the blue limit to exclude
stars near the halo turnoff). Figure 12 plots those data, and shows a clear
discontinuity between at [Fe/H]$\sim -1$. The metal-poor stars span the
magnitude range from V=9 to $\sim15$, with the bulk of the stars fainter than 11th
magnitude, but only 30 of the 332 stars with [Fe/H]$\ge -1$ have
V$>$11.0 mag. This is exactly what one would predict given the disk and halo
kinematics outlined in the previous section:
the relatively bright limiting magnitude of the Hipparcos sample eliminates only
a small fraction of the disk/thick disk stars. The Hipparcos data provide a
fair sample of the F and G stars in the Galactic disk.

The CLLA sample included a total of 541 stars with 0.55 $\le$(B-V)$\le$0.70.
Table 3 lists the apparent magnitude distribution for all those stars, and
for the stars with Hipparcos parallax measurements, in both cases 
dividing the subsamples at [Fe/H]=-1. Almost 90\% of the metal-rich stars with
V$<$10 have Hipparcos data, but coverage falls to only $\sim 35\%$ for the fainter stars.
However, the broad distance distribution described by proper-motion selected
samples mitigates the resultant selection effects. We
have used Monte-Carlo simulations to examine how the increased incompleteness
at fainter magnitudes affects the detection efficiency as a function of
absolute magnitude. We generate a volume-complete sample of stars with $r < 150$
parsecs, +4$\le M_V \le 8$, and kinematics matching those of the thick disk, 
and identify those with proper motions $\mu > 0.25$ arcsec yr$^{-1}$. 
Our model assumes that 75\% of the stars with 
V$<$10 are selected for astrometric observation, but only 30\% of
those with 10$<$V$<$11, and none fainter than V=11. Thus, our calculation 
should provide a conservative estimate of the bias against including
lower-luminosity stars in the Hipparcos observations. Our results show negligible
effects for stars with M$_V \le 5$, while the relative 
detection efficiency drops to 85\% at M$_V = 5.25\pm0.25$; 70 \% at
M$_V = 5.75$; and 50 \% at M$_V = 6.25$. Based on the 47 Tuc colour-magnitude
diagram, we expect thick disk stars with (B-V)$\sim 0.7$ to have absolute magnitude
no fainter than M$_V \sim 6$. Thus, the thick disk should be well-represented within
this colour-selected subset of CLLA sample with Hipparcos data.

\subsection{Thick disk stars in the Lowell proper motion catalogue}

\cite{cll89} have proposed that their analysis of the kinematics
and abundance distribution of stars in the Lowell catalogue provides 
evidence for the presence of a kinematically distinct, [Fe/H]$\sim$-0.5 
stellar population, which they identify with the thick disk. Our Hipparcos
observations provide no new data on the stellar abundance distribution of
the latter stars, but we can test whether their distribution in the 
colour-magnitude diagram matches that predicted on the basis of the
(M$_V$, (B-V), [Fe/H]) calibration derived by \cite{lcl88}. Inverting the
problem, we can use figure 11 to estimate the abundance distribution that
is consistent with the observed displacement below the main-sequence.

There are 248 stars in the CLLA sample which colours in the range 0.55 to 0.70
in (B-V), spectroscopic abundance estimates and Hipparcos parallaxes measured
to a precision of ${\sigma_\pi \over \pi} \le 0.15$. As described in the previous
section, these stars provide a fair, unbiased sample of the disk/thick disk stars
in the Lowell survey. Figure 13a compares the absolute magnitudes estimated by
CLLA for the 192 stars with no evidence of binarity
against those derived for the same stars from the Hipparcos astrometry. Twelve
stars have M$_V^{HIP} < 4.2$ and are probably subgiants, and those stars
are identified separately in the figure. There is a systematic offset in the
sense that the CLLA absolute magnitude estimates are fainter than the directly-determined
M$_V^{HIP}$. The
Hipparcos parallaxes are typically 5 mas (but as much as 13 mas)
smaller than those inferred from the (M$_V$, abundance) calibration.

Figure 13b matches the observed ($\delta M_V$, [Fe/H]$_{CLLA}$) distribution
against the mean calibration derived from figure 11, limiting the comparison 
to the 180 single stars in the sample. The sample is dominated by [Fe/H]$< -1$
disk dwarfs, as expected, but the few halo subdwarfs in the sample do lie
close to the mean relation. At higher abundances, however, the stars tend 
to lie above the calibrating relation - that is,
there are significantly fewer subluminous stars than one would expect given 
the number of intermediate-abundance stars identified by CLLA. 
As a result, if we use our ($\delta M_V$, [Fe/H]) relation to infer the
metallicity distribution of the sample, the number of stars in the low-abundance
([Fe/H]$<-0.5$) tail of the distribution is reduced significantly. 
Based on the (M$_V$, (B$-$V)) distribution, only $\sim 15\%$
of the stars have $-0.3 \le [Fe/H] \le -1.1$, while the spectroscopic calibration 
places approximately 40\% of the stars in that abundance range. Since the relative
number of the higher-velocity (thick disk) component is amplified by a factor of $\sim2.7$
within this proper-motion limited sample (section 5.2), this implies that only
$\sim 5\%$ of a volume-limited sample of local disk stars fall within this metallicity
range.

This result is emphasised in figure 14, where we plot the 
(M$_V$, (B$-$V)) colour-magnitude diagram for all stars in the CLLA sample with
-0.3 $\le [Fe/H] \le -0.9$ and with Hipparcos parallaxes measured to a precision of 15\%
or better. (The data are plotted using Lutz-Kelker corrected M$_V$, but those
corrections amount to 0.12 magnitudes at most.) 
As a reference, we plot colour-magnitude relations for 
NGC 6397, M5 and 47 Tucanae; for the nearby stars; and 
for the two old, open clusters M67 and NGC 6791. The M67 schematic is
derived from Montgomery et al's (1993) CCD photometry of cluster members (their figure
8), adopting their reddening estimate of E$_{B-V}$=0.05 mag. With an age of 
$\sim 5$ Gyrs (\cite{ph94}), M67 has an abundance of [Fe/H] $\sim -0.09$ dex (\cite{fr95}), and
matching the upper main-sequence to the nearby-star relation gives (m-M)$_0$=9.6 mag.
NGC 6791, on the other hand, appears to be metal-rich (\cite{kr95}), despite being one of the
oldest known disk clusters ($\sim 9-10$ Gyrs, \cite{mjp}). The cluster is highly reddened,
with estimates ranging from E$_{B-V}$=0.10 to 0.23 mag (\cite{kr95}). Adopting
Kaluzny \& Rucinski's preferred value of E$_{B-V}$=0.17 mag leads to an estimated
distance modulus of (m-M)$_0$ = 12.8 mag if we match the fiducial sequence (from \cite{mjp})
against the Hyades (M$_V$, (B$-$V)) relation. 

The three sets of panels in figure 14 include 237 of the 441 stars which CLLA 
identify as having abundances between [Fe/H]=$-0.3$ and $-0.9$.
Although the data are subdivided by abundance, there is relatively little difference 
amongst the three subsets in the observed distribution in the 
(M$_V$, (B$-$V)) plane. To underline that point, the right-hand panels
in figure 14 plot the (M$_V$, (B-V)) distribution predicted for these stars
by CLLA. The disparity between the predicted and observed distributions in
the two lower-abundance subsets is particularly evident. A modest fraction 
of those stars are clearly subgiants, but there is also a significant offset
between the mean relation defined by the main sequence stars in each subgroup.
Almost one-third of 
the stars in the -0.5 $>$ [Fe/H]$_{CLLA} >$ -0.9 abundance range are identified
as binaries, and unresolved binaries clearly can skew the observed distribution
to brighter absolute magnitudes. These stars are identified, however, in
figure 14, and, while many lie towards the upper edge of the 
distribution in M$_V$, even allowing for an equal-mass companion generally 
leaves the star closer to the Solar Neighbourhood fiducial than to 
the 47 Tucanae main-sequence. 
It is therefore unlikely that the higher average luminosities can be
ascribed to this cause. Nor, as we have discussed at length, is the 
discrepancy likely to reflect biases in the Hipparcos sample selection.
We therefore conclude that there is a significant mismatch between
the observed and predicted luminosities of the mildly metal-poor stars
in the CLLA proper-motion sample.

\subsection {Discussion}

The results plotted in figure 14 can be summarised very simply: there are
significantly fewer subluminous stars than one would
expect given the spectroscopic abundance distribution derived by CLLA.
We have presented arguments as to why this cannot be 
ascribed solely to selection effects: the subset of the CLLA sample
observed by Hipparcos is a fair sample of the stars in the Lowell survey.
One can explain this result in three ways: first, one can postulate
that the presence of systematic errors in the Hipparcos parallax
measurements. However, those errors must amount to from 4 to 13 mas
to account for the substantial differences between the observed and predicted
colour-magnitude distributions plotted in figure 14. Errors of this
magnitude would have been detected easily in the extensive comparison
with previous ground-based observations undertaken by Arenou et al (1995). 
Indeed, a simple cross-referencing against the full Yale Parallax catalogue 
(van Altena et al, 1995)
would reveal systematics at this level. None such are evident.

The two remaining hypotheses are: that the
CLLA abundance analysis systematically underestimates the metallicity of mildly
metal-poor stars, placing a disproportionate fraction of the sample in
the range $-0.4 \le [Fe/H] \le -1$; or, that the abundance
estimates are accurate, but there is little correlation 
between luminosity and metallicity 
for stars with [Fe/H] $> -1$ dex. Either conclusion has serious implications:
if the latter alternative is correct, then it removes the fundamental
assumption in main-sequence fitting (that local subdwarfs lie on the
same main-sequence as cluster stars of similar mass and abundance), and
casts doubt on the distance moduli derived for metal-rich globulars and metal-poor 
open clusters; if the former conclusion is correct, then this calls into question
the conclusions derived by CLLA from their analysis of the
properties of the higher-abundance (non-halo, thick disk) stars in their
sample. 

We can probe this issue using results from the \cite{s93} uvby
survey of Lowell proper motion stars. \cite{sn89} use
observations of stars with known abundances (based on high-resolution
spectroscopic analyses) to calibrate relations between the Stromgren
photometric indices ((b-y), m$_1$, c$_1$) and [Fe/H]. \cite{cl96} compare
their own abundance determinations to the Stromgren scale, and finding a
systematic offset, in the sense that [Fe/H]$_{SN}$ gives higher abundances 
for mildly metal-poor stars, and an average scatter in the residuals of only $\pm 0.17$ dex.
That comparison, however, is limited to stars from \cite{sn88}, a sample
which consists predominantly of HD stars brighter than 10th magnitude. We have
extended the comparison to include the Lowell proper-motion stars observed
by Schuster et al.

Figure 15a compares Stromgren-based and CLLA abundance measurements for 420 stars
from the Lowell survey. Stars identified as
spectroscopic binaries (either confirmed or suspected) by CLLA, or as having
nearby faint companions, are plotted as open symbols. Figure 15b plots the same diagram, but 
omitting binaries and stars with significant (E$_{B-V} > 0.02$) reddening. We also
distinguish between stars brighter and fainter than V=10. The dispersion of
residuals in figure 15b is $\sigma_{[Fe/H]} \sim \pm 0.2$dex, increasing
to $\sim \pm 0.3$ dex at lower abundances, although note that (as expected) most
of the metal-poor stars have $V>10$. There is a systematic offset of $\sim 0.2$dex,
independent of abundance and magnitude (fig. 15c), in the sense that the
Schuster et al. scale is more metal rich. Figure 15d shows that the residuals
also show a strong correlation with (b-y).

Figure 15 also matches the uvby-abundance determinations against the high-resolution
analyses of the subdwarf stars listed in Table 1. The comparison with the AFG
metallicities shows considerable scatter, perhaps reflecting differences in the
effective temperature scales, and a small systematic offset ($\sim 0.1$ dex)
in the sense that [Fe/H]$_{SN}$ gives lower abundances. The latter offset
is larger ($\sim 0.15$ dex) and more evident in the comparison with the
GCC analyses and may well be tied to the revision in the solar iron abundance.\footnote
{Gratton et al (1997c) have re-calibrated the Schuster \& Nissen uvby scale
using high-resolution data for a large sample of nearby stars.} 

Some 283 of the stars included in the Schuster et al  catalogue have
Hipparcos parallaxes measured to a precision of 15\% or better. However, only
thirty-four of those stars have abundances [Fe/H]$_{SN} < -0.5$. Figure 16 is
analagous to figure 14 in that it compares the observed colour-magnitude
diagrams for the uvby sample, segregated by abundance, against 
fiducial colour-magnitude relations. We also plot subluminosity, $\delta M_V$, 
as a function of [Fe/H]$_{SN}$ for the 109 stars with 0.55 $\le$ (B-V) $\le$ 0.75.
The distribution is more symmetric than that described by the CLLA stars in figure 13.

Combining the offsets between the CLLA and Stromgren, and the Stromgren and
high-resolution abundance scales leads to the conclusion that there is a
a systematic offset of $\sim0.3$ dex, independent of [Fe/H], between the CLLA 
abundance scale and that defined by the high-resolution analyses which mark
the reference scale used in this paper. Clearly, such an
offset can account, at least
partially, for the systematic discrepancy between the predicted and observed
luminosities of the lower-abundance stars in the CLLA sample (figure 14).
Any such systematic change in the abundance scale would have clear implications
for Galactic structure - notably the population structure in the local disk.
\cite{cll89} have argued for the existence of a discrete thick disk component
largely on the basis of a secondary peak in the abundance distribution of
the Lowell stars at [Fe/H]$_{CLLA} \sim -0.4$. That proposition is supported to some
extent by Gilmore et al's (1995) abundance analysis of $\sim 120$ G-dwarfs at
heights of from 1.5 to 3 kpc above the Plane: the metallicity distribution
shows a broad peak centred at [Fe/H] $\sim -0.6$. The latter study does not, however,
address the issue of the separateness of the parent thick disk population. 

The CLLA abundance distribution of the stars plotted in figure 13 does indeed
show a subsidiary peak at [Fe/H]$\sim -0.4$. However, that secondary maximum is
not reflected in the (M$_V$, (B-V)) distribution of those same star. Reversing 
the usual procedure, and estimating [Fe/H] from the HR diagram position
leads to a metallicity distribution centred on [Fe/H]$\sim0$
and with a relatively small tail to lower abundances. This distribution is more
consistent with a continuum than with a superposition of two distinct populations.

Figure 14 implies that stars with -0.9 $\le$ [Fe/H]$_{CLLA}$ $\le$ -0.3 all
describe essentially the same main sequence. As noted above, one can explain 
this result either by postulating a weak correlation between [Fe/H] and M$_V$,
or by a scale error in the abundances. The most effective method of investigating
which alternative is preferred would be to obtain high signal-to-noise, high-resolution
spectra of Lowell stars with Hipparcos data, choosing the sample to span both
a range in [Fe/H] at constant M$_V$ and constant colour (T$_{eff}$) and a
range in M$_V$ at constant [Fe/H] and (B-V). Differential analysis will reveal 
whether the different luminosities are due to metallicity differences
(i.e. miscalibration of [Fe/H]$_{CLLA}$), or to an intrinsic scatter
in the (M$_V$, (B$-$V)) relation for stars with the same atmospheric 
composition. 

The revised distances for the Lowell survey stars derived from the 
Hipparcos astrometry also affect the
kinematics deduced by CLLA for the [Fe/H]$> -1$ stars in their sample.
CLLA derive their space motions from direct measurements of the radial
velocity, and transverse velocities calibrated 
based on their photometric parallax scale (\cite{lcl88}). 
The larger distances that we derive from the Hipparcos parallaxes
lead to inferred tangential velocities that are higher by up to 30 \%.
A full re-analysis of the kinematics is outwith the scope of this paper,
but, combined with the data plotted in figure 13, the results imply
an increased number of high-velocity, near-solar abundance stars. 

Finally, only $\sim 3 \%$ of the 2048 Lowell stars with high-precision
(${\sigma_\pi \over \pi} < 15 \%$) Hipparcos parallaxes are evolved stars,
yet approximately one in five of the stars with -0.9 $\le$ [Fe/H] $< -0.7$
plotted in figure 14 has an absolute magnitude consistent with its being a
subgiant. If those stars have abundances compatible to M71, then the position
of the corresponding turnoff of the parent population is M$_V^{TO} \sim 4.25$,
implying an age of over 14 Gyrs. However, it is also apparent that higher-abundance
stars, either main-sequence binaries or $\sim 5$ Gyr subgiants, can occupy the
same position in the (M$_V$, (B$-$V)) plane. Given the observed dispersion amongst the
low-resolution abundance indicators, it is possible that the evolved stars
are closer to solar abundance, and therefore younger than 10 Gyrs. 
Again, high-resolution
analysis of individual stars is required before broad conclusions can be
drawn concerning the global star formation history of the Galaxy.

\section {Conclusions}

We have used main-sequence fitting techniques to determine the distances to
globular clusters with abundances in the range -0.7 $\le [Fe/H] \le$ -1.8. The
local calibrators are subdwarfs with both accurate metallicities and with accurate Hipparcos 
parallax measurements. Both the subdwarf and globular cluster abundance scales adopted
in this paper are based on conventional analyses of high-resolution spectroscopy.
Main-sequence fitting gives results which are in general agreement 
with the larger distances, and consequent brighter turnoff luminosities, derived in
both our previous analysis (\cite{re97}) and by \cite{g97}. Using models calculated by
\cite{da97} to estimate absolute ages, these data suggest that the time interval between
the onset of star formation in the halo and subsequent star formation in the disk was 
only 1-2 Gyrs. We identify 47 Tucanae and M71 as being significantly younger than the average 
age of the clusters included in the present sample. 

The \cite{cg97} abundance scale places M5 at an abundance of -1.1 dex. As a result, 
there is a conflict between the absolute magnitudes inferred for the cluster
RR Lyraes and those derived based on statistical parallax analysis of the field
variable in the Solar Neighbourhood. In addition, an M$_V$ of +0.5 magnitude
for [Fe/H]$\sim -1.1$ RR Lyraes implies a distance of $\sim 9.3$ kpc to the
Galactic Centre, a value over 2$\sigma$ from the currently-preferred value
of 8.0 kpc (\cite{re93}). One can account for the discrepancy, however, if the
cluster horizontal branch stars have a higher helium abundance in the envelope, as
would be expected if dredge-up is responsible for the high [Al/Fe] ratios observed
in the more luminous cluster red giants (Sweigart, 1997). To date, such abundance
anomalies have not been detected in field halo RGB stars (Shetrone, 1997).

We have examined the (M$_V$, (B$-$V)) distribution of Lowell proper-motion stars 
with both accurate Hipparcos parallax measurements and with spectroscopic metallicity
estimates by \cite{lcl88}. Our results show that there are significantly fewer subluminous
stars than one would expect given the [Fe/H]$_{CLLA}$ distribution. Matching the latter
abundance estimates against Stromgren photometry suggests that there is a systematic
bias in the CLLA scale such that metallicities are underestimated. If we use the cluster
(M$_V$, (B$-$V)) sequences to calibrate the absolute-magnitude/metallicity distribution
of G and early-K dwarfs, then we find that less than 15\% of the proper motion stars have
abundances below -0.3 dex, half the number with spectroscopic abundance estimates in that
same range. We argue that the metallicity distribution inferred from $\delta M_V$, the
displacement below the nearby-star main-sequence, is more consistent with a 
continuum (the disk) than with a combination of two separate populations, the old disk
and thick disk.

Finally, we should emphasise that, in contrast to pre-Hipparcos studies, 
uncertainties in the absolute magnitudes of the calibrating
subdwarfs {\sl do not} dominate the error budget in the 
cluster distance determinations in the current analysis.
There are sufficient stars with high-precision parallax measurements
to render the results independent of whether systematic
Lutz-Kelker corrections are applied or not. The major source of
uncertainty in main-sequence fitting now rests in the colour, rather
than absolute magnitude, domaine.
The line-of-sight reddening to each cluster; the (T$_{eff}$, 
colour) calibration of the theoretical isochrones, and hence the 
mono-metallicity $\delta (B$-$V) / \delta ([Fe/H])$ corrections; and
the metallicity scales adopted for both the clusters and the field
subdwarfs, all affect colours (specifically (B$-$V)) to a greater 
extent they affect than magnitude. The steep slope of the main-sequence guarantees that 
any systematic error in colour is amplified by 
a factor of five in distance modulus. The most problematic of these 
uncertainties is the abundance calibration - a matter underscored by both the
contrast between the expected  and observed colour-magnitude distribution of
what are nominally mildly metal-poor field stars, and the significant scatter
in the comparison between low-resolution abundance estimators. 
Further observations designed to
determine the source of this discrepancy, and verify the abundance scale
at intermediate abundances, are clearly of the highest priority.

\acknowledgments 
{This research made use of the Simbad database, operated at CDS, Strasbourg,
France. INR was supported partially by NSF grants AST-9412463 and AST-9318984.
Hipparcos data for the subdwarf stars were obtained as part of joint proposals
with G. F. Gilmore. The author also acknowledges useful comments by R. Peterson and
by the referee, B. Carney.}

\clearpage
\centerline{FIGURE CAPTIONS}
\vskip2em

\figcaption{The formal uncertainties in the Hipparcos parallax measurements, plotted as
a function of apparent magnitude.}

\figcaption{ Monte-Carlo simulations of Lutz-Kelker corrections. {\sl A:} The bias in M$_V$ as
a function of parallax precision for a uniformly-distributed sample with r$<$ 1000 pc,
M$_V$=5 and $\sigma_\pi = 1$ mas. The solid and dotted lines show the Hanson (1979)
n=4 (uniform density) and n=3 approximations. {\sl B:} the same simulation, but with a magnitude limit
added at V=11.5. {\sl C:} the effect of allowing $\sigma_\pi$ to vary with apparent magnitude,
as in figure 1. {\sl D:} a dispersion of $\sigma_M = 0.25$ mag has been included. We plot the predicted 
bias for M$_V$=+6 (triangles) and M$_V$=+3 (open circles) besides the M$_V$=+5 (crosses)
prediction}

\figcaption{A comparison between the abundances, gravities and temperatures calculated by
Gratton et al (1997a) and Axer et al (1994) from analysis of high-resolution spectroscopic data.}

\figcaption{The Carney et al (1994) abundance determinations compared against metallicities
derived by Axer et al (1994) and Gratton et al (1997a). }

\figcaption{a: The (M$_V$, (B$-$V)) colour-magnitude diagram described by the stars listed in Table 1.
The smaller solid dots mark the position of stars within 25 parsecs of the Sun. The solid line
marks a nearby-star main-sequence relation derived by Reid \& Murray (1993), and the dotted line
outlines the Pleiades main-sequence relation for (m-M)=5.3. No Lutz-Kelker corrections have
been applied to the subdwarf data.
b: The two fiducial sequences plotted in figure 5a compared to the (M$_V$,(B-V))
distribution of Hipparcos stars with $\pi > 30$ mas and with ${\sigma_\pi \over \pi} < 0.08$.
c: the abundance distribution of the calibrating subdwarfs. The five binaries excluded
from the main-sequence fitting analysis are marked as open squares.}

\figcaption{ Stars classified as dwarfs and subgiants in the high-resolution analyses. In each
case, stars with log(g) $<$ 4.3 are plotted as filled symbols. It is clear that the
Axer et al (1994) 
analysis misclassifies a significant number of {\sl bona-fide} dwarfs as
subgiants}

\figcaption {The correlation between subluminosity, $\delta M_V$, the distance that a
star falls below the Reid \& Murray (1993) nearby-star relation, and abundance for stars 
with 0.55 $< (B$-$V) <$ 0.75. The dotted line shows the dependence predicted by the
D'Antona et al (1997) isochrones at (B$-$V)=0.60. }

\figcaption { Main-sequence fitting distance determinations for the five clusters
listed in Table 2. In each case the observed parameters for the subdwarfs are
plotted as open squares, while the corrected positions (used in the fitting)
are plotted as solid squares. The solid line marks the Reid \& Murray (1993) nearby-star
relation.}


\figcaption {The observed main-sequence turnoff absolute magnitudes plotted against
the predictions of the D'Antona et al (1997) models. Since the models include no
[$\alpha$/Fe] enhancement, the cluster abundances have been adjusted
by +0.2 for [Fe/H] $<$ -1.2 and by +0.15 at higher abundance (Salaris et al, 1993).}

\figcaption {A comparison between the star-formation history of the Galaxy deduced
by Chaboyer et al (1996), based on globular cluster distances derived using the cited
horizontal branch (M$_V$, [Fe/H]) relation, and the halo cluster ages deduced from
figure 11. The crosses plot the ages of Galactic open clusters (from Friel (1995)),
while the box (WD) marks the age estimated for the Galactic disk from analysis
of the white dwarf luminosity function (Oswalt et al (1996)).}

\figcaption {The dependence of main-sequence luminosity on abundance defined by the calibrated
sequences of the globular clusters 47 Tucanae, M5 and NGC 6397. As in
figure 8, the dotted line shows the dependence predicted by the D'Antona et al (1997)
models at (B$-$V)=0.6, while the empirical relations are plotted for three
colours.}

\figcaption{The apparent magnitude distribution of stars in the full CLLA sample
with colours in the range 0.55 $\le$ (B-V) $\le$ 0.70. Note the distinctly different
magnitude distributions of the disk stars, at [Fe/H]$\le -1$, and the 
higher-velocity dispersion, metal-poor halo subdwarfs.}

\figcaption{A comparison between the predicted and observed absolute magnitudes
for 225 single stars with 0.55 $\le (B-V) \le 0.70$, and both CLLA abundance data
and Hipparcos parallaxes measured to a precision of ${\sigma_\pi \over \pi} \le 0.15$.
The top panel plots, as a function of [Fe/H]$_{CLLA}$, the difference between 
the CLLA predicted M$_V$ and the Hipparcos measurements. Probable subgiants 
are plotted as open triangles. The middle panel matches $\delta M_V$ (from
M$_V$(Hip)) against the mean relation from figure 11. The final panel
compares the abundance distribution that we infer from the single-star 
$\delta M_V$ distribution (excluding subgiants)
against that derived directly from the CLLA spectroscopy.}

\figcaption{The (M$_V$, (B$-$V)) colour magnitude diagrams described by stars from the CLLA
sample with both spectroscopic abundances in the range -0.3 to -0.9 dex and Hipparcos
parallaxes with ${\sigma_\pi \over \pi} \le 0.15$. The data are subdivided into three
abundance subgroups. In each case, the left hand panel plots the Hipparcos-calibrated
absolute magnitudes, while the right-hand panel plots the predicted distribution from CLLA.
The Solar Neighbourhood main-sequence is shown as a solid line, and 
the three globular cluster sequences plotted are for 47 Tucanae (solid line); M5 (dotted line); and
NGC 6397 (dashed line). The M67 ([Fe/H]=-0.09, 5 Gyrs) and NGC 6791 ([Fe/H]=+0.2, 
9-10 Gyrs) colour-magnitude relations are also plotted for the higher-abundance
stars. The encircled points denote stars identified by CLLA as definite or 
suspected spectroscopic binaries}

\figcaption{a: A comparison between the abundances derived by Schuster et al (1993) for 
Lowell proper-motion stars and the CLLA metallicities. Open triangles mark
stars identified as definite or possible spectroscopic binaries (CLLA) or stars
with nearby faint companions which might affect the Stromgren colours; 
b) as in a), but omitting all binaries and stars with E$_{B-} > 0.02$ mag.
Open squares denote stars brighter than V=9 magnitude. The subsequent four
plots are restricted to these stars.
c) the residuals plotted
as a function of apparent magnitude - most of the highly-discrepant points lie
at V $>$ 9th magnitude; d) residuals as a function of (b-y) - it is clear
that there is a correlation with colour; e) the abundance comparison
between the Stromgren photometric calibration and the Gratton et al (1997)
high-resolution analyses; f) the Schuster et al abundances compared to the
metallicities derived by Axer et al (1994).}

\figcaption {The (M$_V$, (B-V)) distribution of the 283 Schuster et al stars with
high-precision parallaxes. The three stars plotted as crosses in the
lowest-abundance subsample (-0.5 $>$ [Fe/H] $> -1$) have abundances [Fe/H] $< -0.7$.
We also plot the ($\delta M_V$, [Fe/H]) distribution for the 109 stars
with 0.55 $\le$ (B-V) $\le$ 0.75.}

\setcounter{figure}{0}
\begin{figure}
\plotone{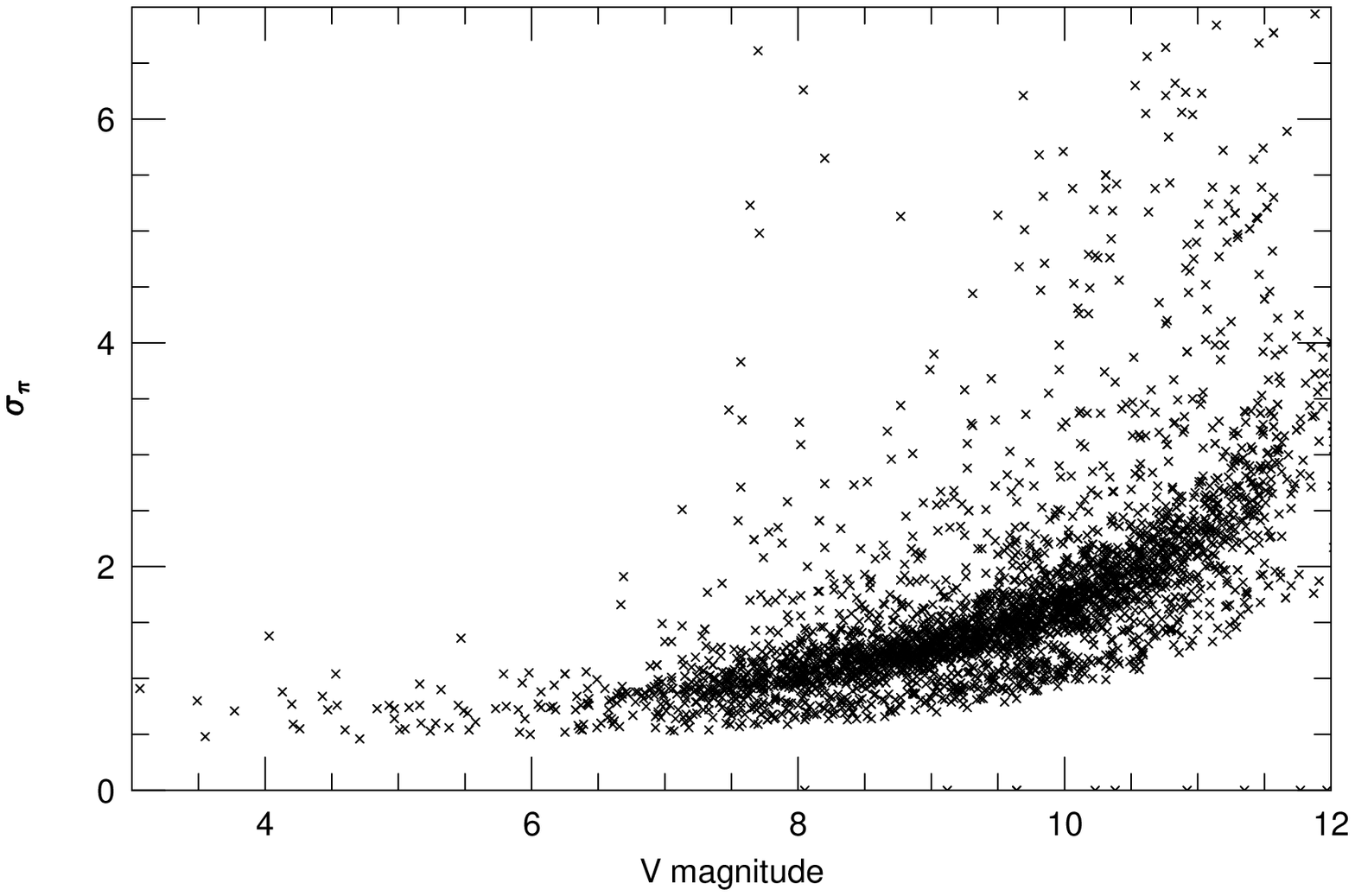}
\caption{}
\end{figure}

\begin{figure}
\plotone{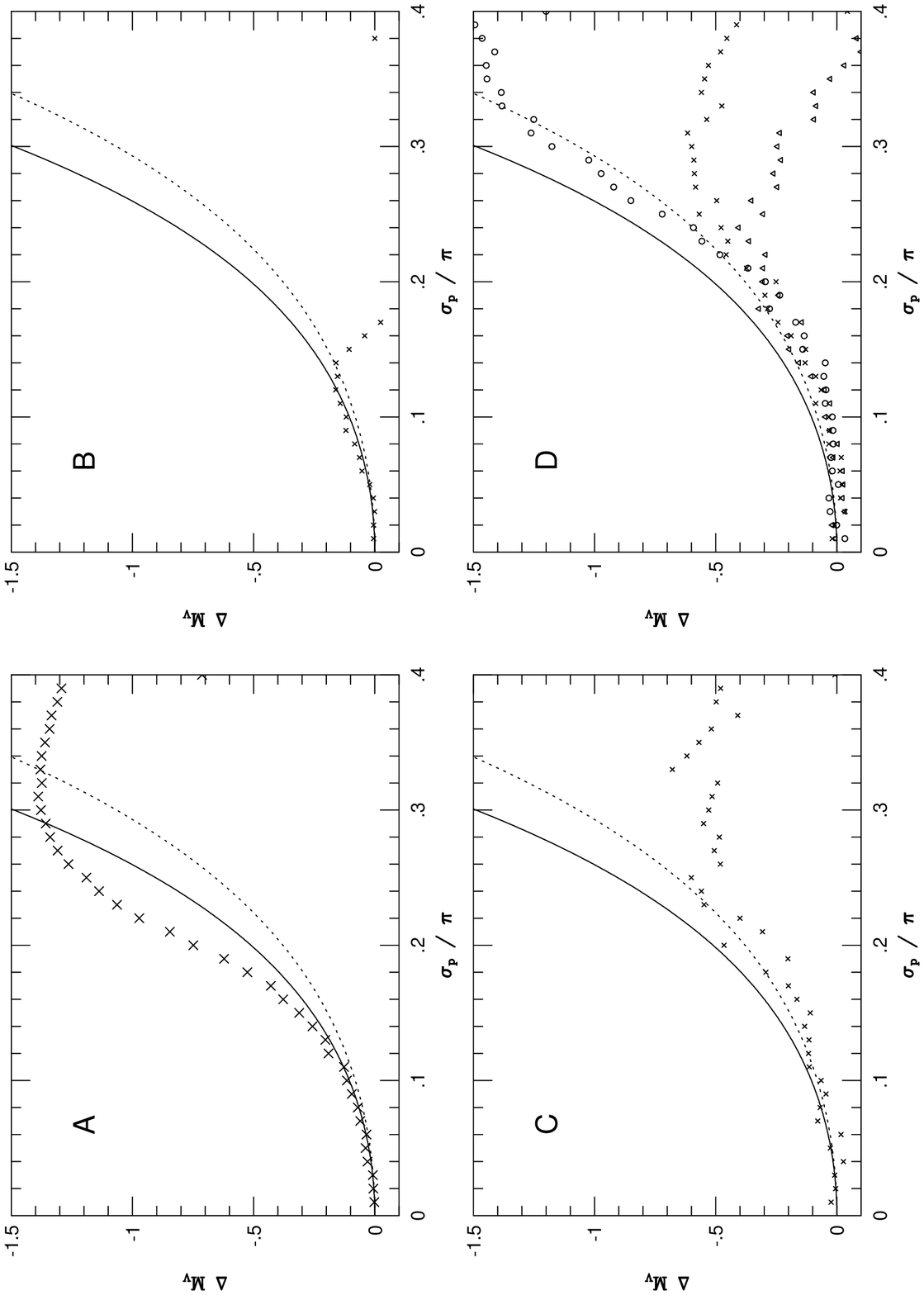}
\caption{}
\end{figure}

\begin{figure}
\plotone{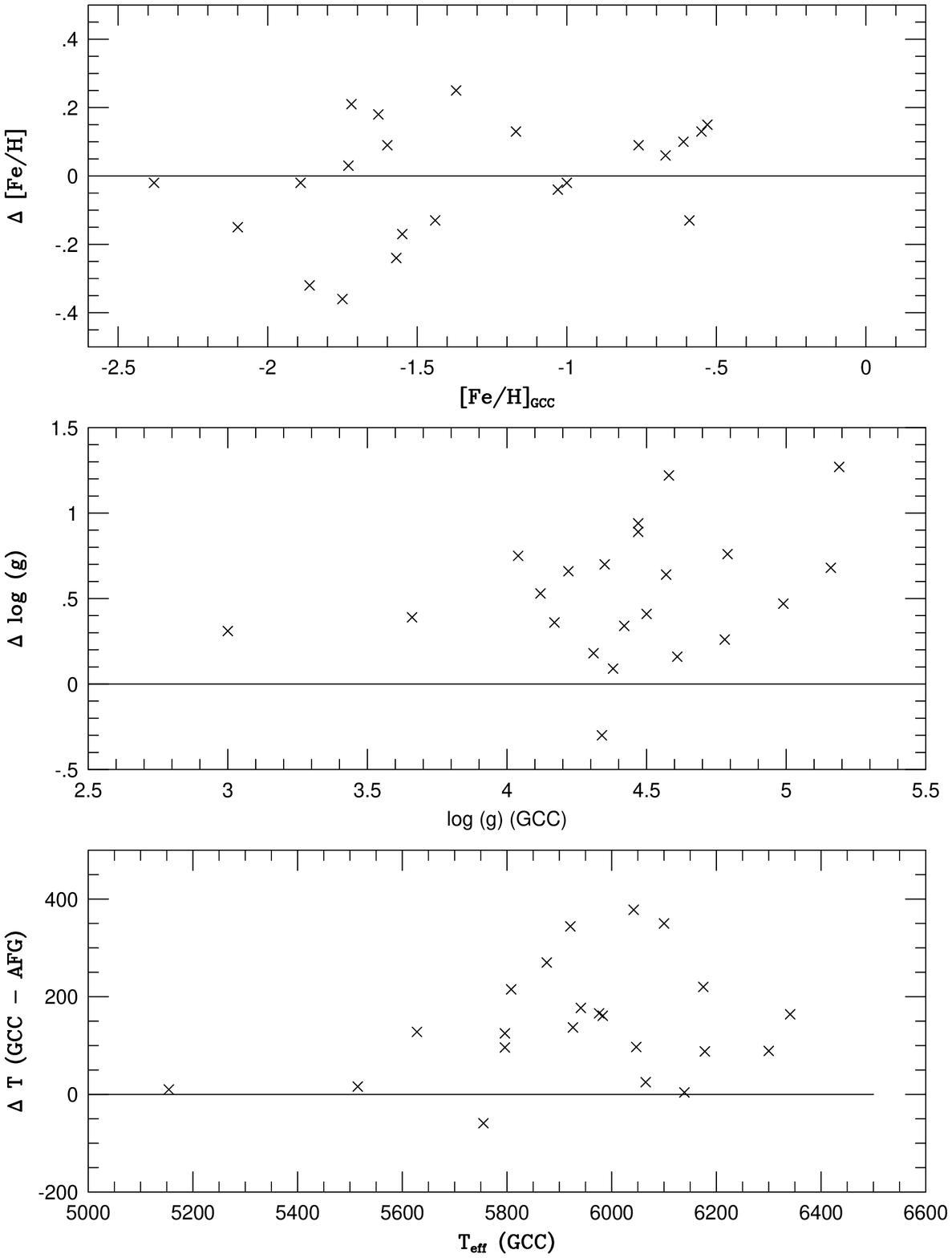}
\caption{}
\end{figure}

\begin{figure}
\plotone{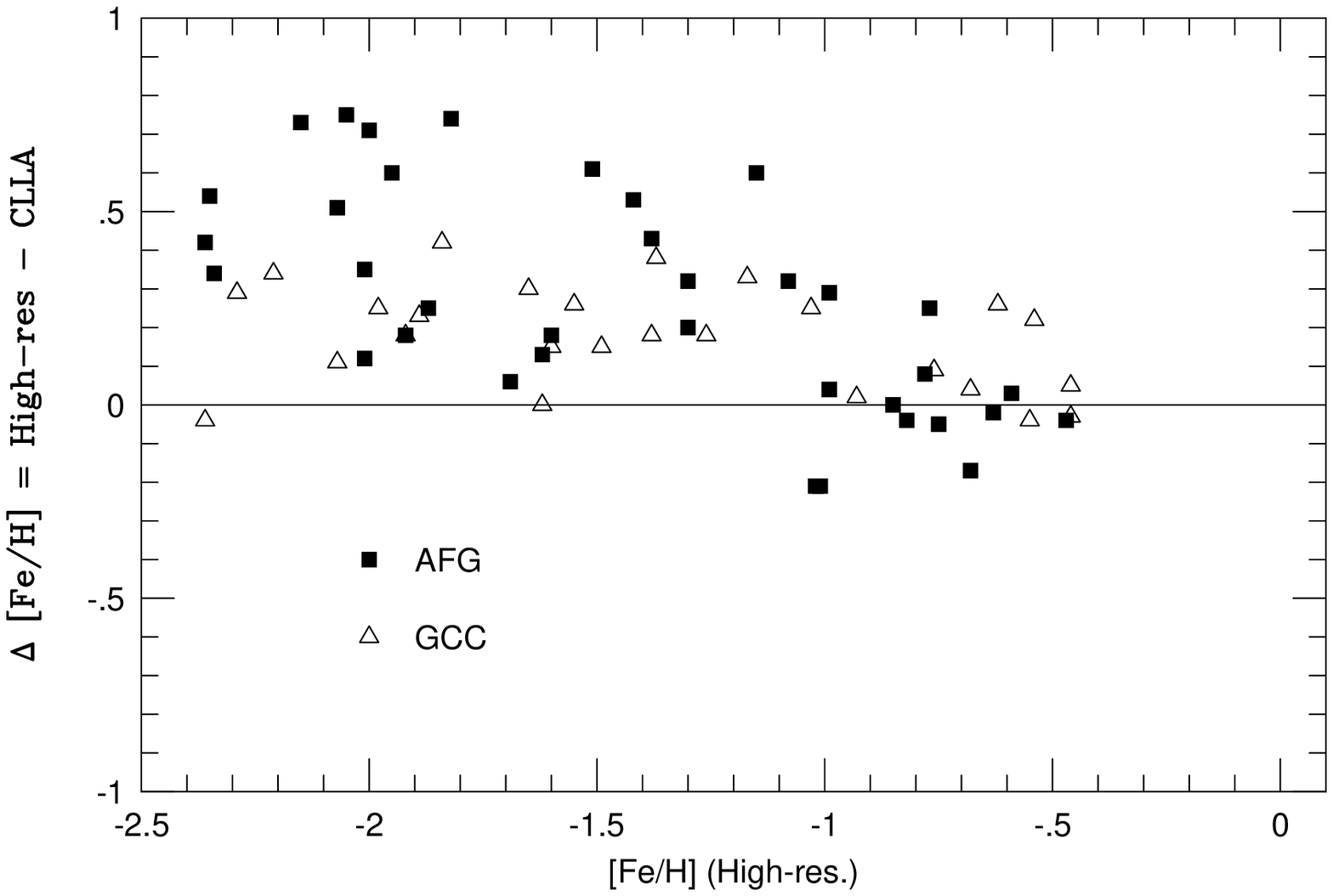}
\caption{}
\end{figure}

\begin{figure}
\plotone{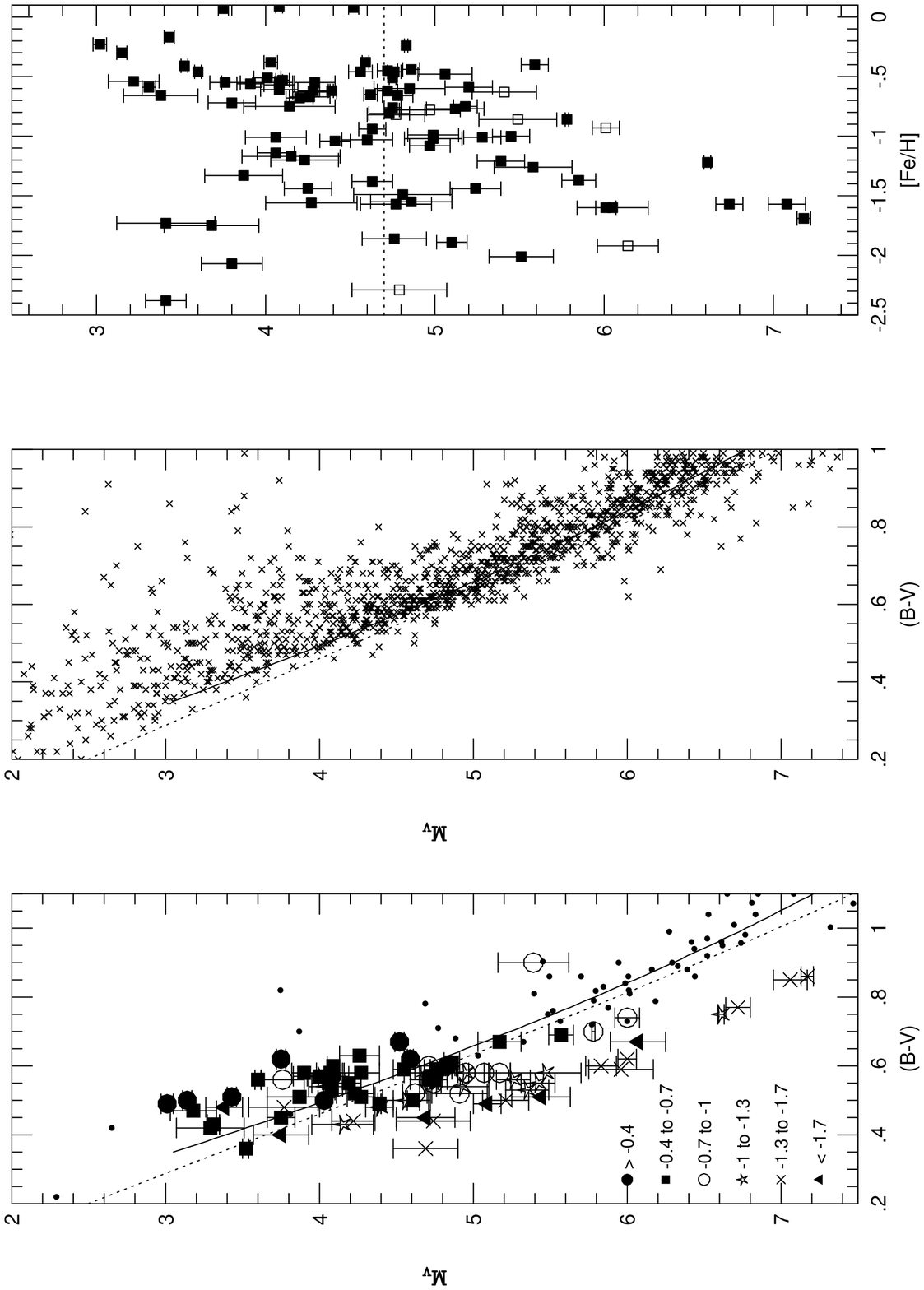}
\caption{}
\end{figure}

\begin{figure}
\plotone{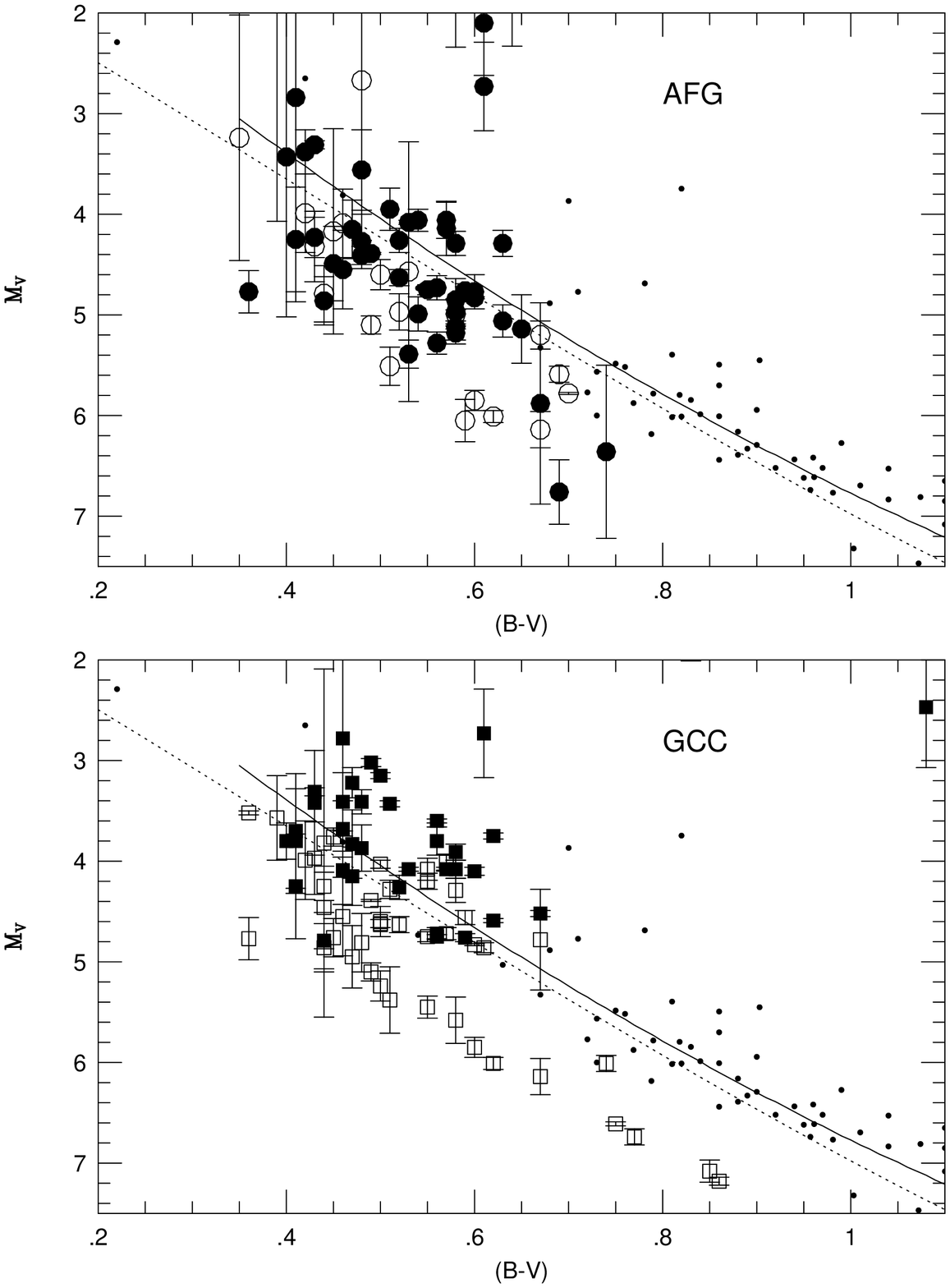}
\caption{}
\end{figure}

\begin{figure}
\plotone{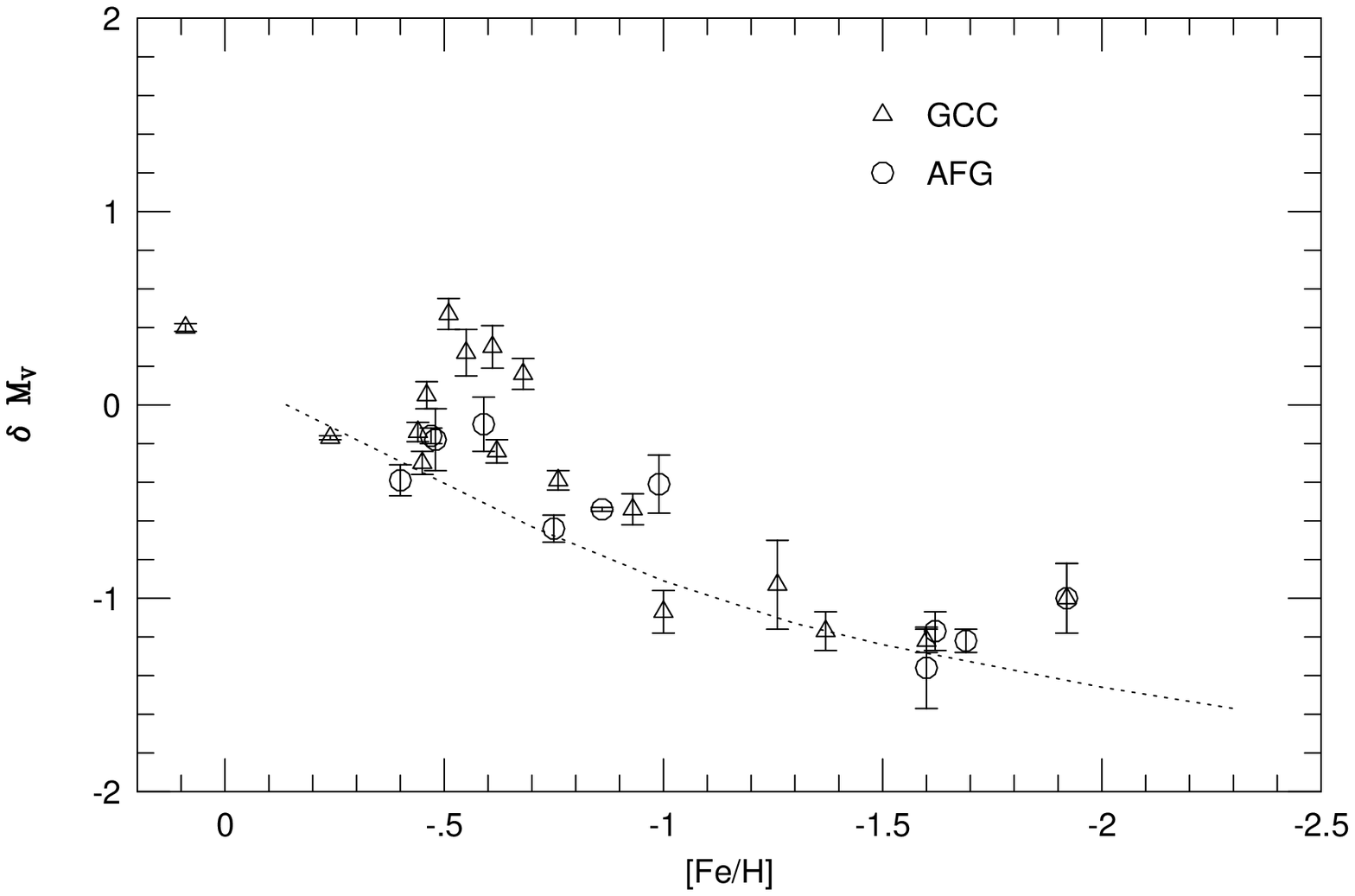}
\caption{}
\end{figure}

\begin{figure}
\plotone{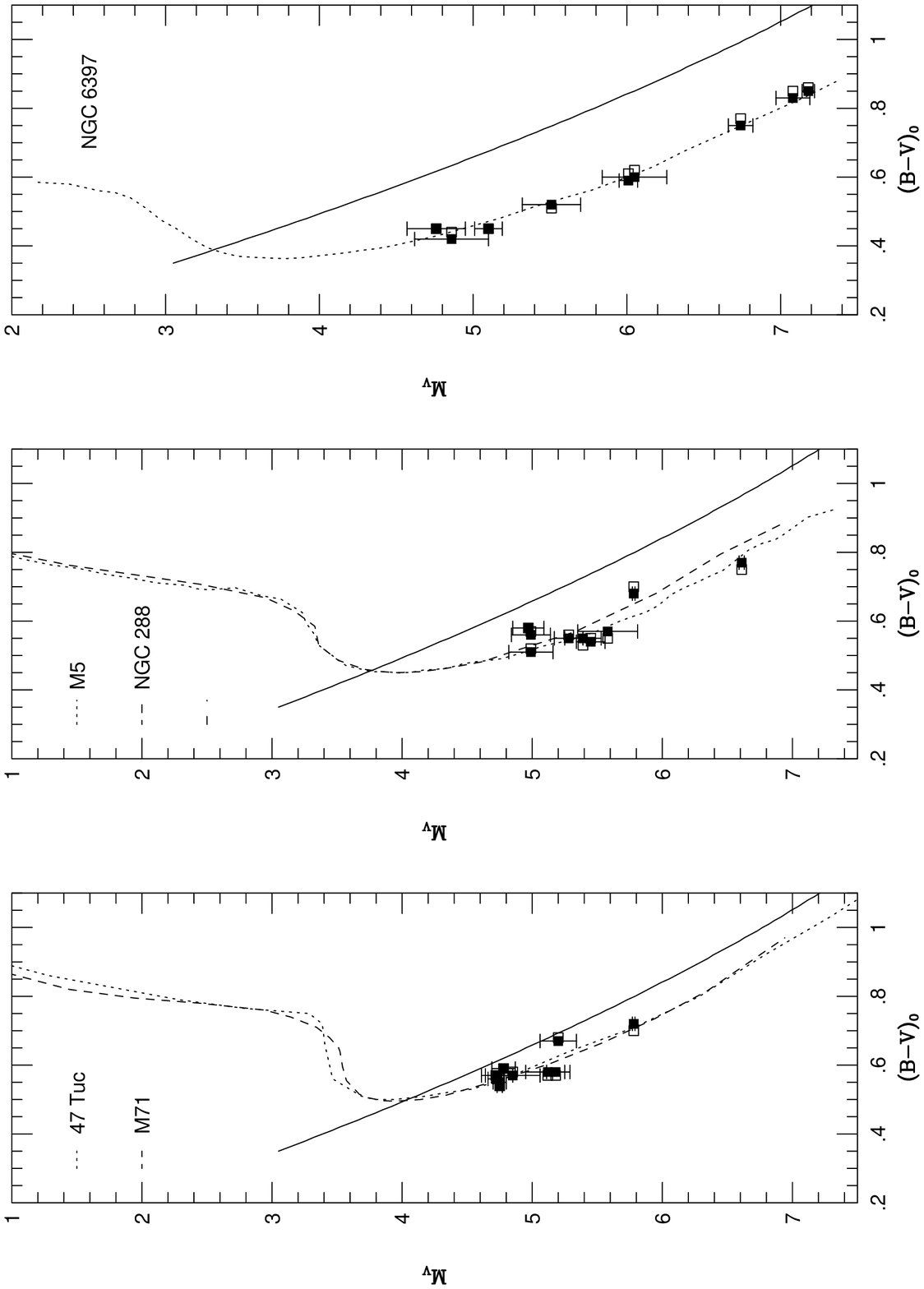}
\caption{}
\end{figure}

\begin{figure}
\plotone{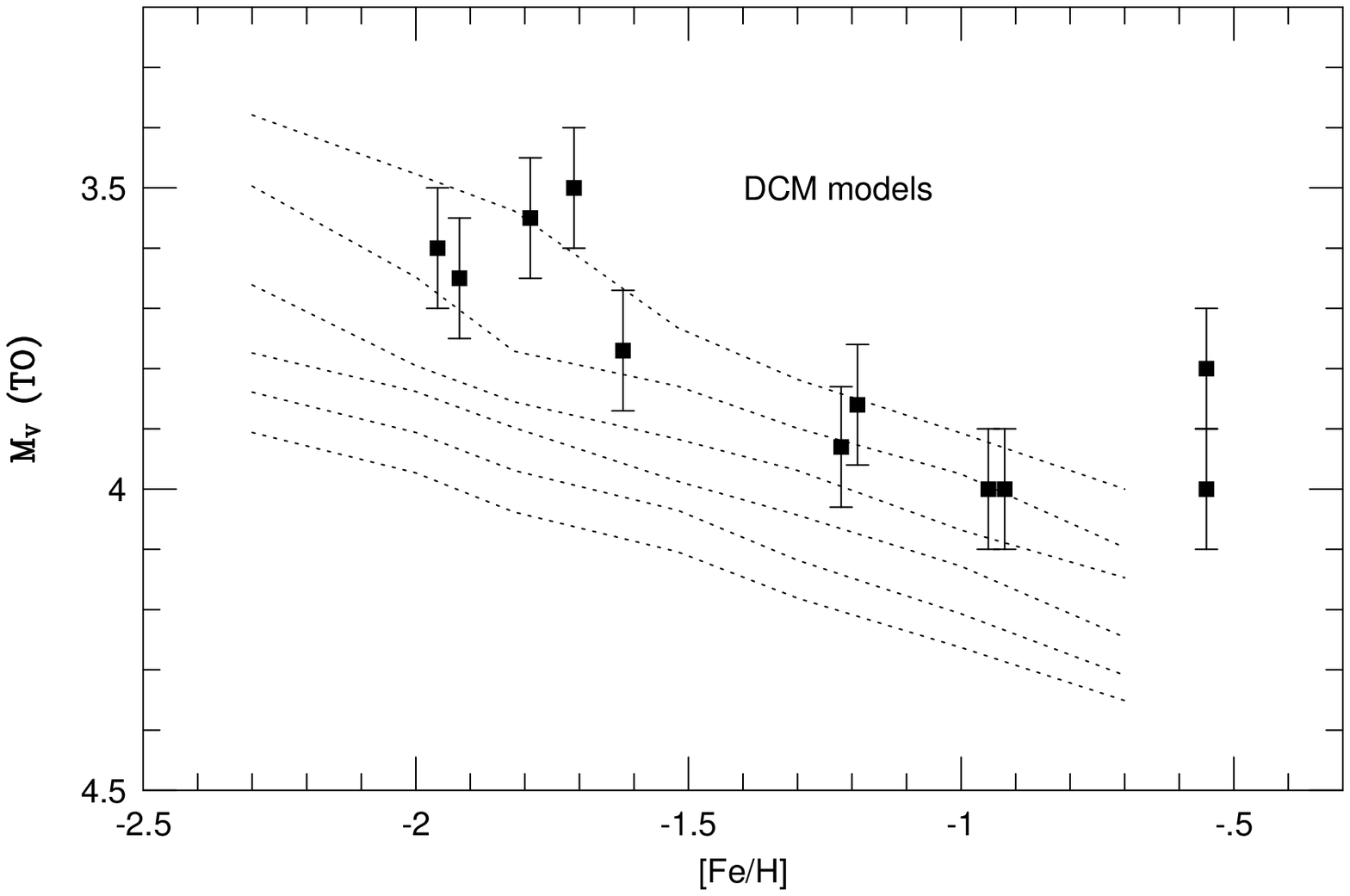}
\caption{}
\end{figure}

\begin{figure}
\plotone{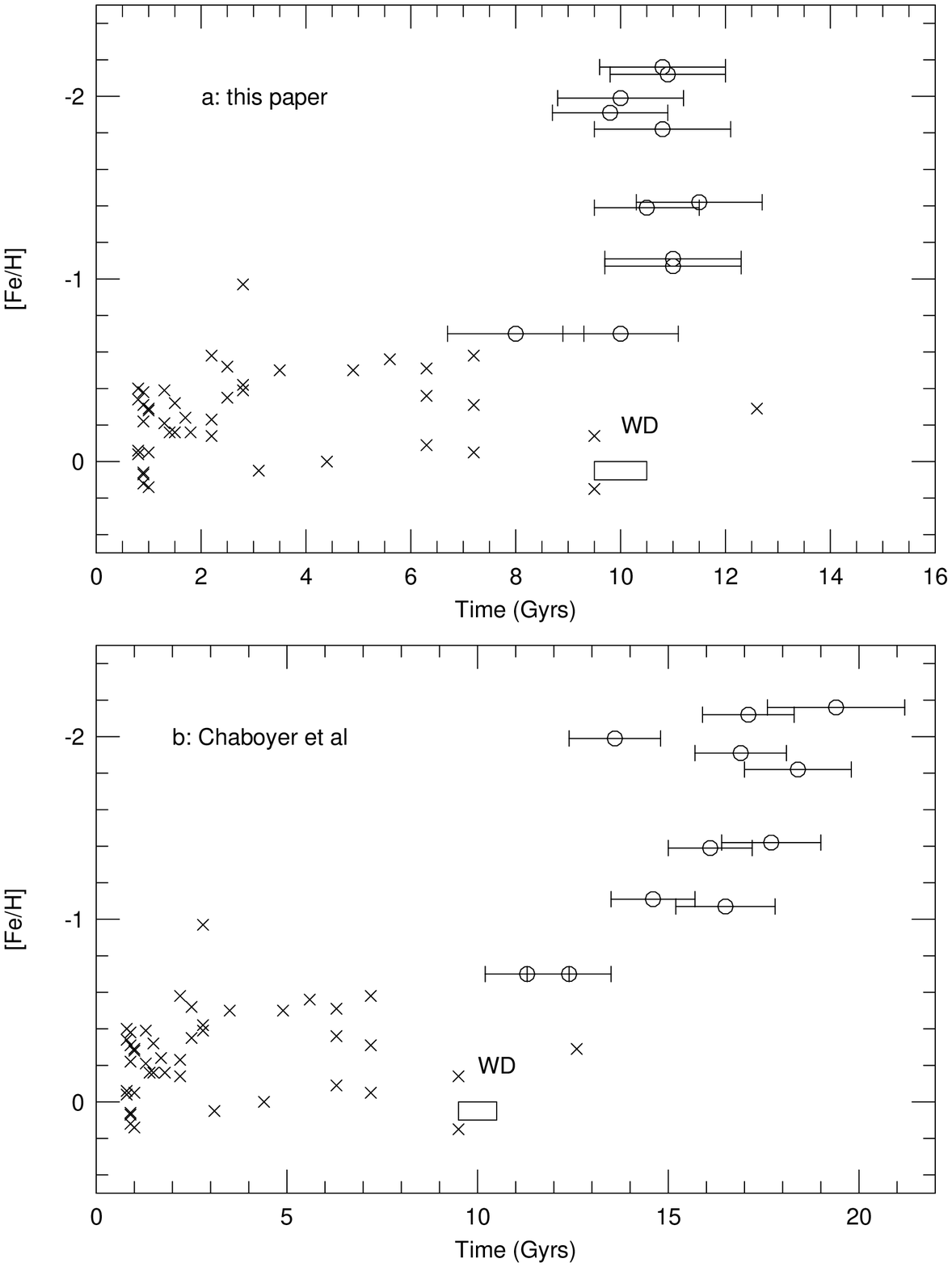}
\caption{}
\end{figure}

\begin{figure}
\plotone{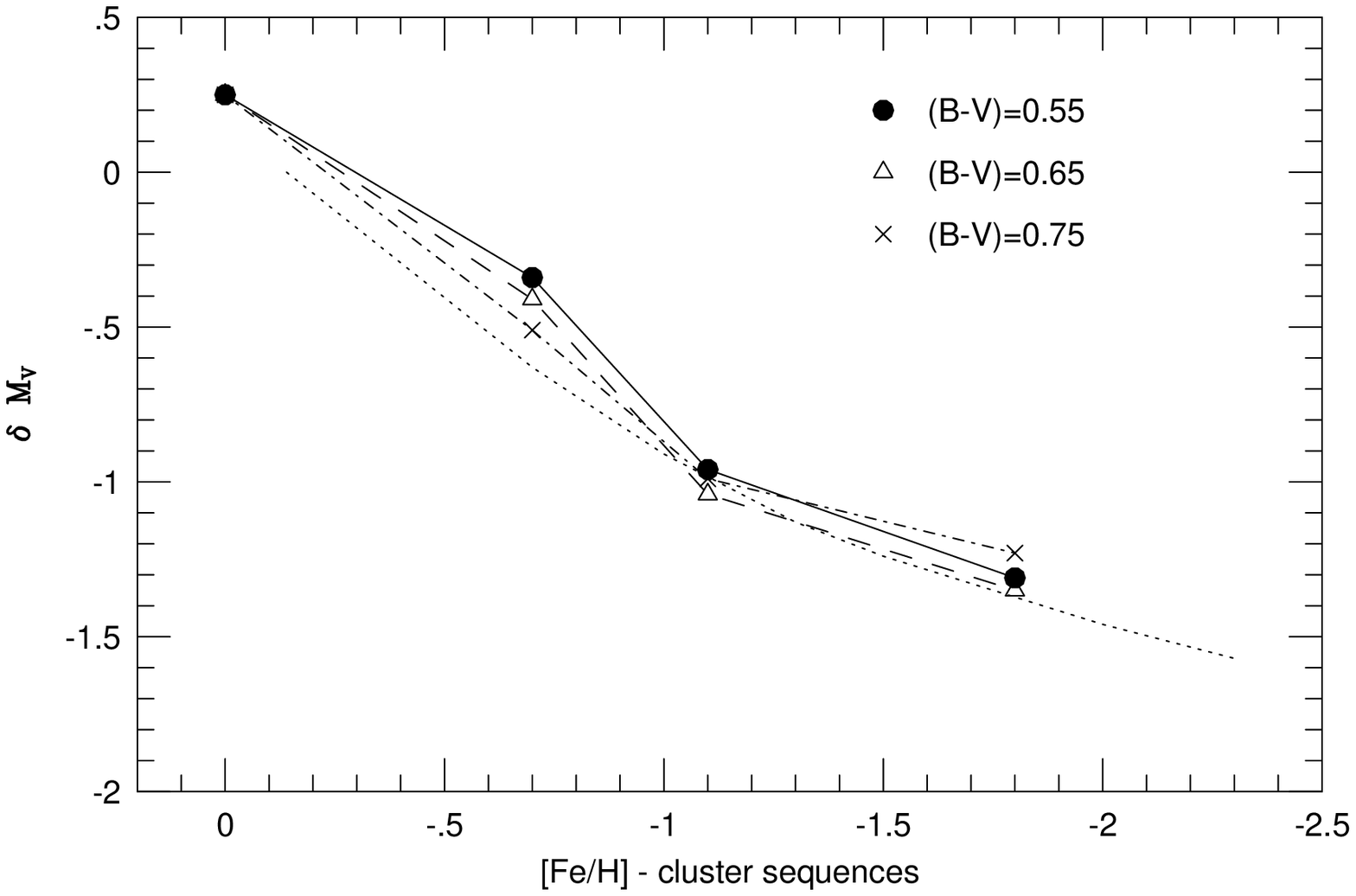}
\caption{}
\end{figure}

\begin{figure}
\plotone{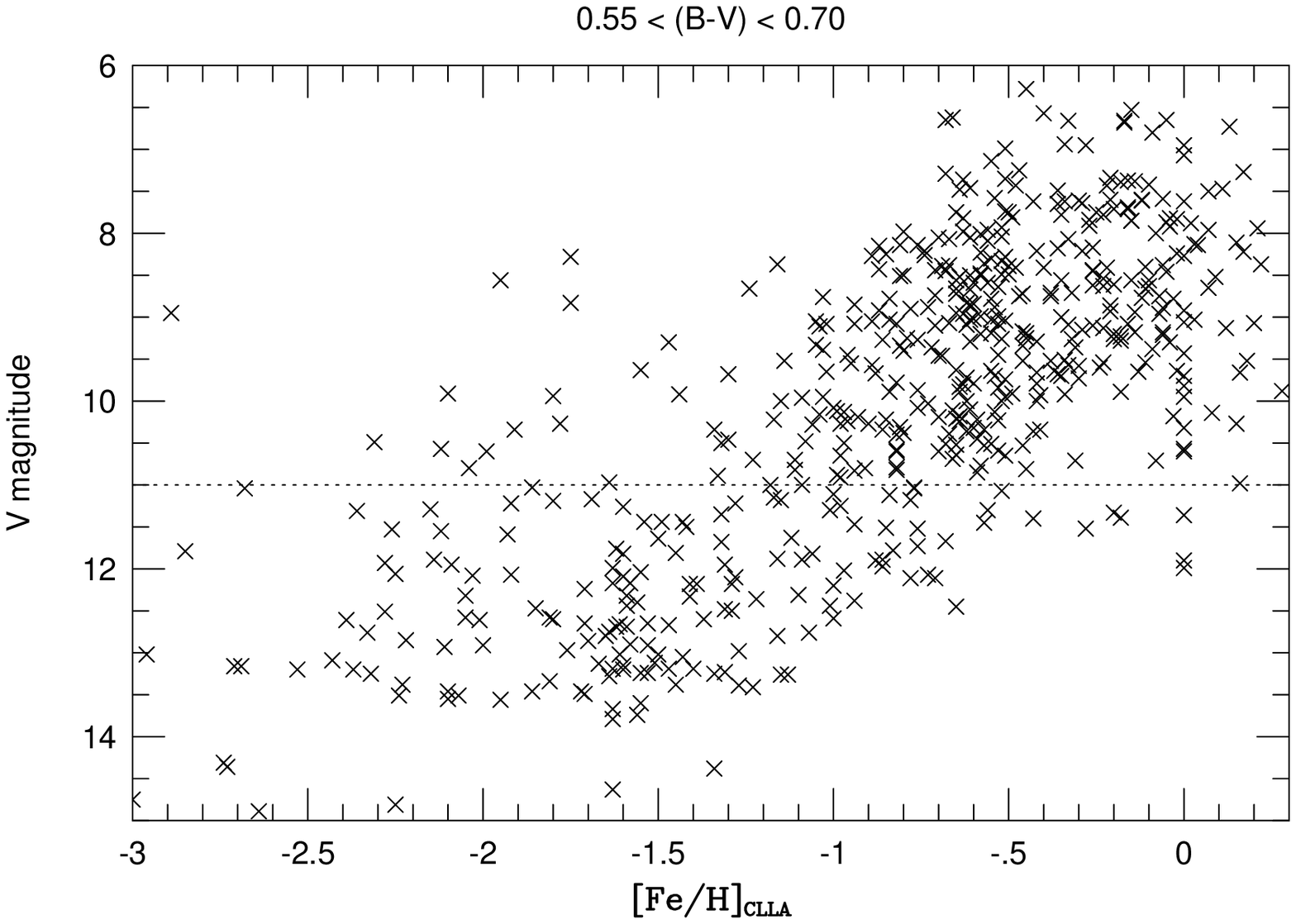}
\caption{}
\end{figure}

\begin{figure}
\plotone{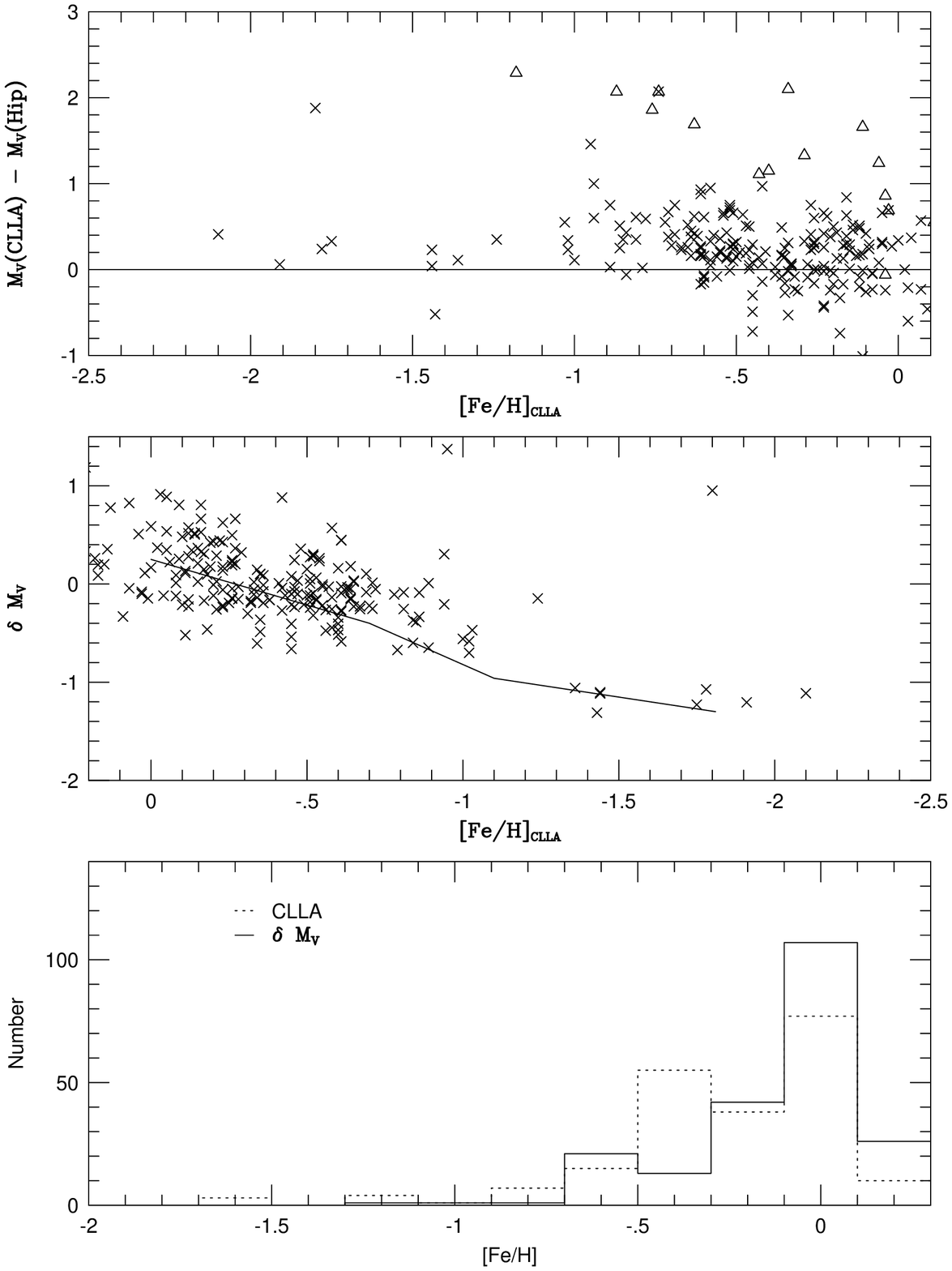}
\caption{}
\end{figure}

\begin{figure}
\plotone{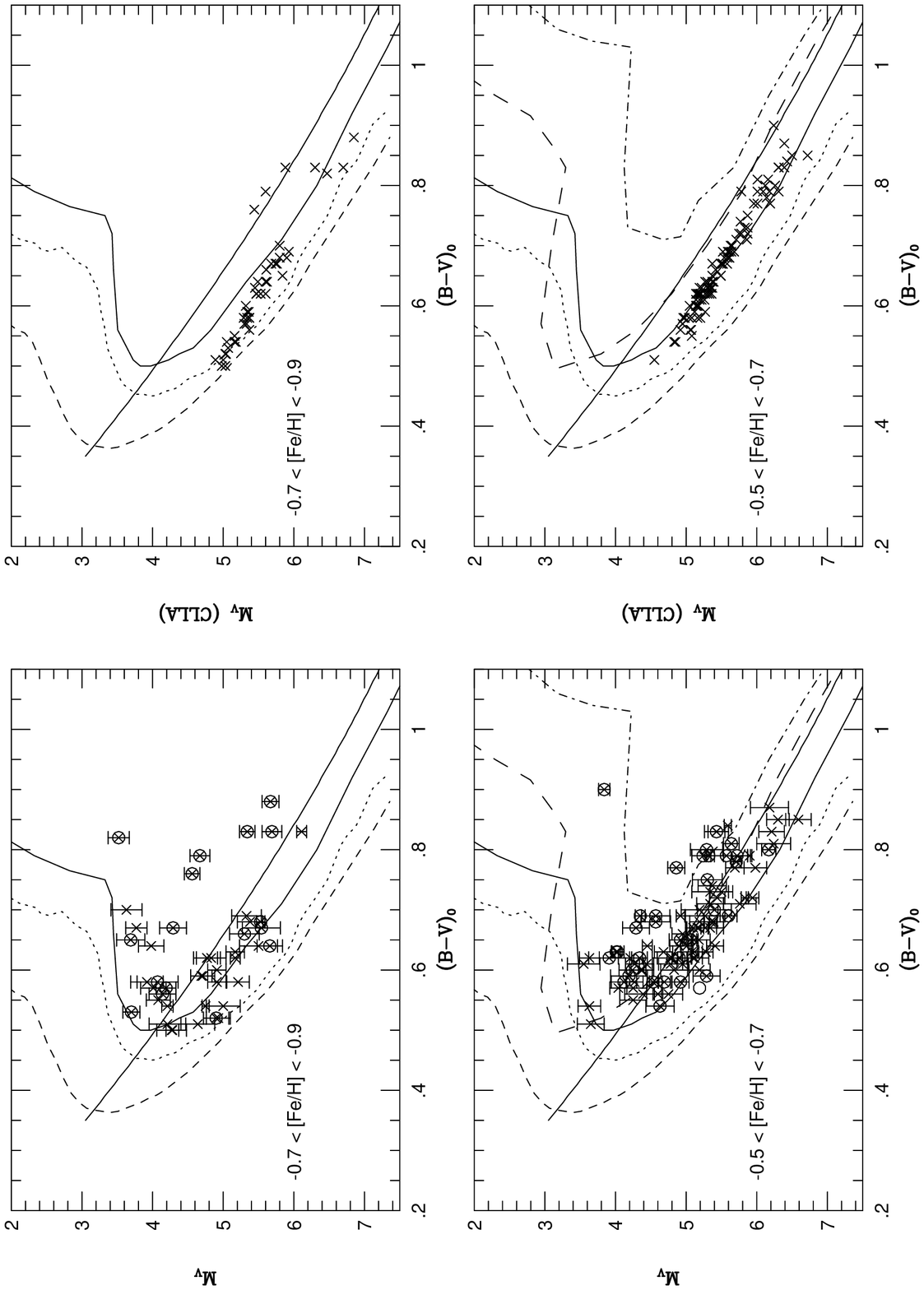}
\caption{}
\end{figure}

\setcounter{figure}{13}
\begin{figure}
\plotone{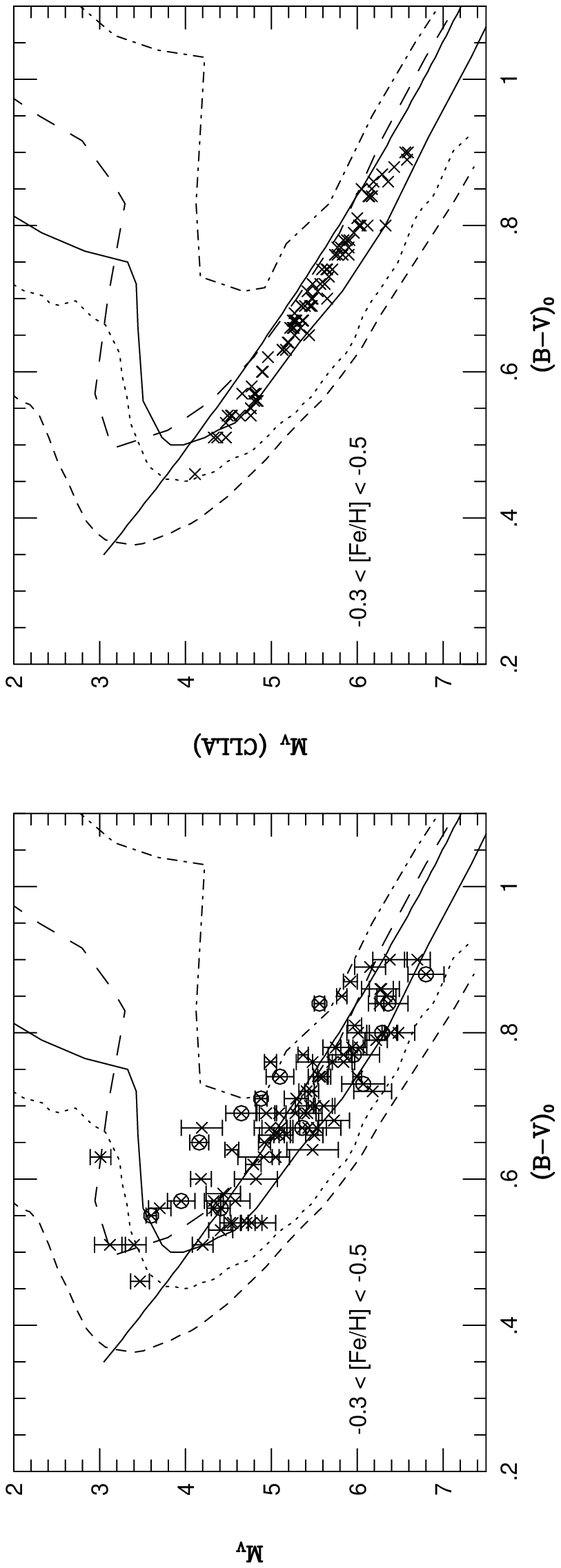}
\caption{}
\end{figure}

\begin{figure}
\plotone{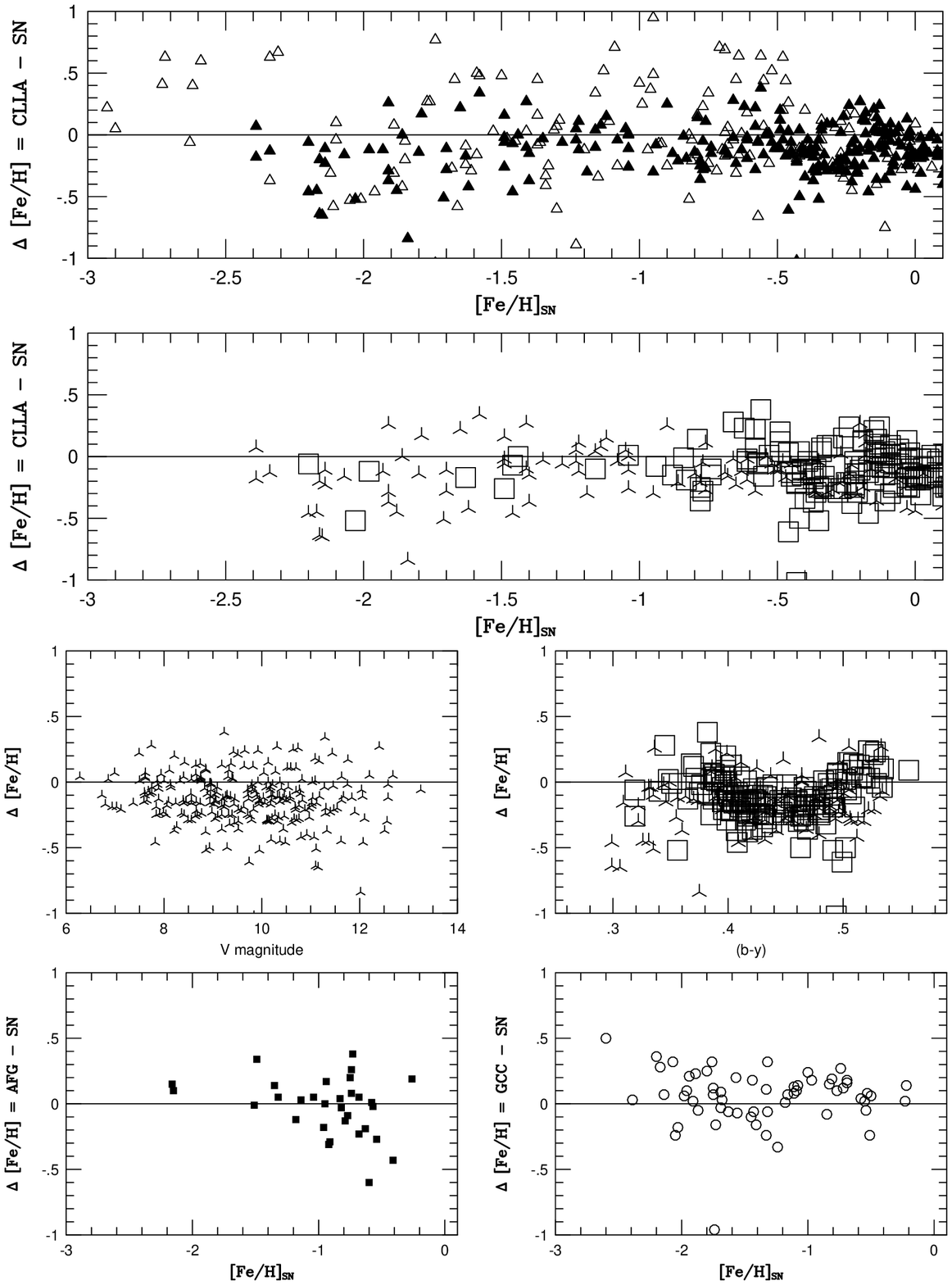}
\caption{}
\end{figure}

\begin{figure}
\plotone{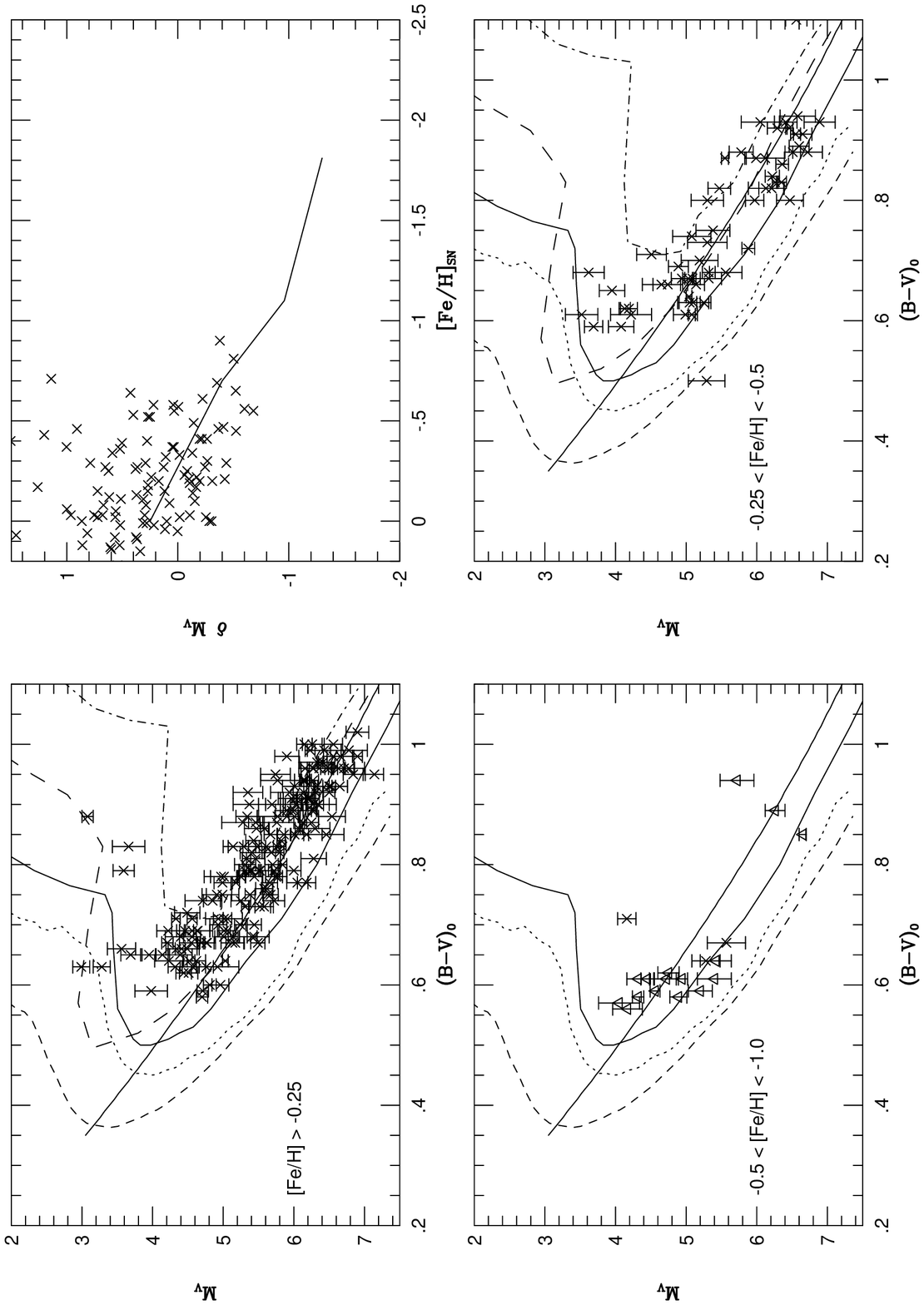}
\caption{}
\end{figure}

\end{document}


\pagestyle{empty}

\begin{deluxetable}{lrrrrrrllllr}
\tablecaption{The local calibrators}

\tablewidth{0pt}
\tablenum{1}
\tablehead{
\colhead{HIC} & \colhead{ V$_0$} & \colhead{(B-V)$_0$} &
\colhead {M$_V$} & \colhead {M$_V^{LK}$ } & 
\colhead{ $\sigma_\pi \over \pi$} &
\colhead{ [Fe/H]} & \colhead { Source} &
\colhead {name} & \colhead {}}
\startdata
    2319&   7.62&   0.47&   3.22&   3.18&   0.070& -0.54& GCC & HD 2615               \nl
    3026&   9.25&   0.46&   4.15&   3.97&   0.144& -1.17& GCC &  HD 3567 &G270-23    \nl
    5336$^1$&   5.17&   0.70&   5.78&   5.78&   0.005& -0.86& AFG &  HD 6582 & $\mu$ Cas  \nl
    5862&   4.97&   0.57&   4.08&   4.08&   0.010&  0.09& GCC & HR 370            \nl
    6159$^2$&   8.90&   0.59&   4.77&   4.72&   0.082& -0.82& AFG &  HD 7983 & G271-34    \nl
    7869$^3$&   8.33&   0.54&   4.06&   4.04&   0.053& -1.14& AFG &  HD 10607 & LHS 1281  \nl
        &  12.44&   1.21&   8.17&   8.15&   0.053& -1.14& AFG &   LHS 1279 & cpm to HD 10607  \nl
   10140$^4$&   8.76&   0.57&   4.99&   4.95&   0.073& -0.99& AFG &  BD+29 366 & G74-5   \nl
   10449&   9.08&   0.57&   5.12&   5.07&   0.083& -0.77& AFG &  BD-01 306 & G159-50  \nl
   12579$^5$&   9.16&   0.52&   4.97&   4.91&   0.088& -0.78& AFG & BD+46 610 & G78-1    \nl
   14594&   8.04&   0.45&   5.10&   5.09&   0.044& -1.89& GCC & HD 19445 & G37-26   \nl
   15371$^6$&   5.24&   0.60&   4.83&   4.83&   0.006& -0.24& GCC & HR 1010 & $\zeta^2$ Ret  \nl
   16404$^7$&   9.91&   0.65&   6.14&   6.08&   0.087& -1.92& GCC & BD+66 0268& G246-38 \nl
   17147&   6.68&   0.54&   4.75&   4.75&   0.021& -0.76& GCC & HD 22879 & G80-15     \nl
   18915&   8.51&   0.86&   7.18&   7.18&   0.020& -1.69& GCC & HD 25329            \nl
   19797&   9.23&   0.36&   4.77&   4.68&   0.104& -1.57& GCC & HD 284248 &G8-16 \nl
   22596$^8$&   6.94&   0.56&   4.56&   4.55&   0.033& -0.46& GCC & HD 30649            \nl
   24316&   9.43&   0.50&   5.24&   5.20&   0.069& -1.44& GCC & HD 34328            \nl
   31188&   8.60&   0.56&   4.73&   4.70&   0.058& -0.81& AFG & HD 46341             \nl
   33221&   9.07&   0.48&   3.87&   3.77&   0.111& -1.33& GCC & CD-33 3337            \nl
   36491&   8.48&   0.52&   4.99&   4.94&   0.083& -1.02& AFG &  HD 59374 & G88-31    \nl
   38541$^9$&   8.27&   0.61&   6.01&   6.00&   0.029& -1.60& GCC &  HD 64090 & G90-25  \nl
   38625$^{10}$&   7.43&   0.73&   6.01&   6.00&   0.036& -0.93& GCC & HD 64606 & G112-54\nl
   40613&   7.74&   0.58&   4.29&   4.26&   0.057& -0.55& GCC & HD 69611 & G113-24    \nl
   40778&   9.73&   0.48&   4.81&   4.64&   0.142& -1.49& GCC & BD+54 1216 & G194-22  \nl
   42734&   7.42&   0.58&   4.08&   4.07&   0.044& -0.57& GCC & HD 74011              \nl
   44811&   7.72&   0.57&   4.72&   4.71&   0.027& -0.62& GCC & HD 78747              \nl
   47139&   8.33&   1.01&  -0.21&  -5.06&   0.500& -1.51& GCC & HD 83212            \nl
   48113&   5.08&   0.62&   3.75&   3.75&   0.014&  0.07& GCC & HR 3881            \nl
   48152$^a$&   8.33&   0.40&   3.80&   3.74&   0.085& -2.07& GCC & HD 84937 & G43-3\nl
   49793&   8.07&   0.59&   4.78&   4.77&   0.041& -0.66& AFG &  HD 88261             \nl
   50384$^{11}$&   5.81&   0.50&   4.03&   4.03&   0.017& -0.38& GCC & HR 4039               \nl
   51415&   9.04&   0.56&   5.28&   5.26&   0.053& -1.01& AFG &  HD 91345             \nl
   51700&   7.50&   0.56&   4.72&   4.71&   0.030& -0.45& GCC & HD 91347            \nl
   53070$^a$&   8.21&   0.47&   4.63&   4.60&   0.059& -1.38& GCC &  HD 94028 & G58-25  \nl
   54641&   8.16&   0.48&   4.41&   4.40&   0.043& -1.04& AFG &  HD 97320             \nl
   55790&   9.07&   0.48&   4.27&   4.12&   0.135& -1.56& AFG &  HD 99383             \nl
   56837&   8.45&   0.69&   5.59&   5.58&   0.040& -0.40& AFG &  BD+45 1949 & G122-39  \nl
   57360&   8.75&   0.43&   4.23&   4.16&   0.096& -1.20& AFG &  HD 102200            \nl
   57450&   9.91&   0.55&   5.58&   5.48&   0.113& -1.26& GCC & BD+51 1696 & G176-53  \nl
   57939&   6.42&   0.75&   6.61&   6.61&   0.007& -1.22& GCC & HD 103095 & G122-51 \nl
   59699&   9.28&   0.57&   4.14&   4.00&   0.130& -0.75& AFG &  HD 106411            \nl
   60551$^a$&   8.03&   0.57&   5.18&   5.17&   0.030& -0.75& AFG &  BD+39 2519 & G123-2\nl
\tablebreak
   60632&   9.66&   0.44&   4.86&   4.75&   0.118& -1.55& GCC & HD 108177 & G13-35   \nl
   60779&   9.28&   0.58&   4.85&   4.76&   0.103& -0.60& AFG &  HD 108405            \nl
   62207&   5.95&   0.56&   4.75&   4.75&   0.011& -0.51& GCC & HR 4845            \nl
   62923$^{12}$&   8.02&   0.51&   3.95&   3.87&   0.102& -0.55& AFG &  HD 111971    \nl
   64426&   7.30&   0.52&   4.26&   4.23&   0.058& -0.67& GCC & HD 114762            \nl
   65201$^{13}$&   8.80&   0.45&   4.76&   4.69&   0.093& -1.86& GCC & HD 116064     \nl
   66509&   8.81&   0.68&   5.20&   5.17&   0.064& -0.59& AFG & HD 118659 & G63-44    \nl
   68246&   8.61&   0.42&   3.38&   3.29&   0.107& -0.66& AFG & HD 121387             \nl
   68464&   8.73&   0.46&   3.68&   3.53&   0.135& -1.75& GCC & HD 122196            \nl
   70520&   7.27&   0.58&   3.91&   3.90&   0.039& -0.56& GCC & HD 126512            \nl
   71284$^{14}$&   4.47&   0.36&   3.52&   3.52&   0.011& -0.41& GCC & HR 5447            \nl
   72461$^{15}$&   9.73&   0.43&   4.79&   4.63&   0.138& -2.29& GCC &BD+26 2606 & G166-45\nl
   72673&   7.16&   0.45&   3.76&   3.75&   0.042& -0.55& GCC & HD 130551            \nl
   74079&   7.67&   0.56&   3.80&   3.77&   0.066& -0.72& GCC & HD 134169            \nl
   74234$^{16}$&   9.44&   0.85&   7.08&   7.06&   0.050& -1.57& GCC & HD 134440            \nl
   74235$^{16}$&   9.07&   0.77&   6.74&   6.73&   0.040& -1.57& GCC & HD 134439            \nl
   76976&   7.20&   0.48&   3.41&   3.39&   0.056& -2.38& GCC & HD 140283            \nl
   77760&   4.60&   0.56&   3.60&   3.60&   0.009& -0.46& GCC & HR 5914            \nl
   80587&   7.69&   0.55&   4.08&   4.06&   0.052& -0.61& GCC & HD 148211            \nl
   80837&   7.27&   0.54&   4.20&   4.19&   0.037& -0.68& GCC & HD 148816 & G17-21   \nl
   84862$^{17}$&   5.38&   0.62&   4.59&   4.59&   0.008& -0.38& GCC & HR 6458            \nl
   84905&   6.95&   0.57&   4.01&   4.00&   0.037& -0.51& GCC & HD 157089            \nl
   85373$^{18}$&   9.67&   0.82&   5.41&   5.35&   0.089& -0.63& AFG &  HD 158226B           \nl
   85378$^{18}$&   8.48&   0.61&   4.29&   4.26&   0.064& -0.63& AFG &  HD 158226A & G181-47 \nl
   86013$^{19}$&   8.37&   0.58&   4.97&   4.95&   0.056& -1.08& AFG &  BD+06 3455 & G139-48 \nl
   86694&   8.72&   0.46&   3.41&   3.23&   0.144& -1.73& GCC & HD 160617            \nl
   88622&   6.80&   0.61&   4.86&   4.86&   0.021& -0.44& GCC & HD 165401            \nl
   88745$^{20}$&   5.05&   0.53&   4.08&   4.08&   0.009& -0.54& GCC & HR 6775 & HD 165908   \nl
   89554&   8.22&   0.44&   4.25&   4.22&   0.065& -1.44& GCC & HD 166913            \nl
   92532&   7.15&   0.54&   4.76&   4.76&   0.018& -0.46& GCC & HD 174912 & G207-5   \nl
   96185&   6.62&   0.60&   4.10&   4.10&   0.020& -0.53& GCC & HD 184499            \nl
   98020&   8.83&   0.59&   5.85&   5.83&   0.046& -1.37& GCC & HD 188510 & G143-17  \nl
   99267&  10.11&   0.51&   5.51&   5.44&   0.094& -2.01& AFG & BD+42 3607 & G125-64 \nl
  100568&   8.65&   0.55&   5.45&   5.43&   0.054& -1.00& GCC & HD 193901            \nl
  100792&   8.33&   0.49&   4.60&   4.56&   0.069& -1.03& GCC & HD 194598 & G24-15   \nl
  102862&   8.94&   0.63&   5.06&   5.01&   0.077& -0.48& AFG &  HD 198245            \nl
  103269&  10.28&   0.62&   6.05&   5.96&   0.103& -1.60& AFG &  BD+41 3931 & G212-7  \nl
  104659&   7.37&   0.52&   4.63&   4.62&   0.036& -0.94& GCC & HD 201891            \nl
  105858&   4.21&   0.49&   4.39&   4.39&   0.005& -0.62& GCC & HR 8181 & HD 203608  \nl
  105888&   8.49&   0.57&   4.06&   4.00&   0.085& -1.01& AFG &  BD+04 4674 & G25-49  \nl
  106749&   9.04&   0.53&   5.39&   5.36&   0.066& -1.21& AFG &  HD 205650            \nl
  107975&   5.52&   0.43&   3.31&   3.31&   0.019& -0.59& GCC & HR 8354 & HD 207978   \nl
  108490$^{21}$&   6.95&   0.50&   4.62&   4.62&   0.021& -0.65& GCC & HD 208906            \nl
  112229&   7.41&   0.50&   4.28&   4.27&   0.041& -0.62& GCC & HD 215257            \nl
  112447$^{22}$&   4.20&   0.50&   3.15&   3.15&   0.013& -0.30& GCC & HR 8665            \nl
\tablebreak
  112935&   5.16&   0.49&   3.02&   3.02&   0.020& -0.23& GCC & HR 8697            \nl
  113357&   5.45&   0.67&   4.52&   4.52&   0.012&  0.08& GCC & HR 8729            \nl
  116771$^{23}$&   4.13&   0.51&   3.43&   3.43&   0.012& -0.17& GCC & HR 8969            \nl
\enddata
\tablenotetext{1} {CCDM J0108+5455, $V \sim 11$ M-dwarf companion}
\tablenotetext{2} {CCDM J01191-0856, m$_1$=9.3, m$_2$=10.8, omitted from MS fitting}
\tablenotetext{3} {CCDM J01412-6741, companion is LHS 1279, $\Delta m$ = 4 mag}
\tablenotetext{4} {CCDM J02104+2948, $\Delta m$ = 4 mag}
\tablenotetext{5} {SB1 CLLA, omitted from MSF}
\tablenotetext{6} {CCDM J03180-6232, wide cpm companion of $\zeta^1$ Ret}
\tablenotetext{7} {Suspected as not single from Hipparcos astrometry, omitted from MS fitting}
\tablenotetext{8} {CCDM J04518+4550, optical companion?, $\Delta m$ = 5 mag}
\tablenotetext{9} {CCDM J07536+3037, SB1? CLLA, optical comp. $\Delta m$ = 5 mag}
\tablenotetext{10} {CCDM J07546-0215, SB1 CLLA, omitted from MS fitting}
\tablenotetext{11} {CCDM J10172+2306, Gl 387AB, $\Delta m$ = 5 mag}
\tablenotetext{12} {CCDM J12590-0950, m$_1$=8.7, m$_2$=9.3. omitted from MS fitting}
\tablenotetext{13} {CCDM J13217-0950, no data on companion}
\tablenotetext{14} {CCDM J14347+2945, two optical(?) companions, both $\Delta m >$ 5 mag}
\tablenotetext{15} {SB1 CLLA, omitted from MS fitting}
\tablenotetext{16} {CCDM J15102-1624, wide well-resolved binary}
\tablenotetext{17} {CCDM J17206+3229, two optical(?) companions, both $\Delta m >$ 4.5 mag}
\tablenotetext{18} {CCDM J17267+3104, wide well-resolved binary}
\tablenotetext{19} {CCDM J17348+0601, ADS 10638C, wide well-resolved binary}
\tablenotetext{20} {CCDM J18071+3034, Gl 704, $\Delta m =$ 3.5 mag}
\tablenotetext{21} {CCDM J21547+2949, optical(?) companion, $\Delta m >$ 6.5 mag}
\tablenotetext{22} {CCDM J22467+1211, optical(?) companions, $\Delta m >$ 7 mag}
\tablenotetext{23} {CCDM J23399+0538, optical(?) companion, $\Delta m >$ 8 mag}
\tablenotetext{a} {Suspected single-lined binary, CLLA. Omitted from MS fitting}
\end{deluxetable}